\documentclass[twocolumn,prd,superscriptaddress,showpacs,floatfix,preprintnumbers, nofootinbib,hyperref]{revtex4}  

\usepackage{graphicx}
\usepackage{epsfig}   
\usepackage{bm}  
\usepackage{color}  
\usepackage[colorlinks]{hyperref}
\usepackage{array}
\usepackage{amsmath}
\usepackage{amssymb}
\usepackage{subfigure}

\providecommand{\apjs}{Astrophys. J. Suppl.}
\providecommand{\aap}{Astron. Astrophys.}
\providecommand{\epja}{Eur. Phys. J. A.}
\providecommand{\physrep}{Phys. Rep.}
\providecommand{\npa}{Nucl. Phys. A}
\providecommand{\npb}{Nucl. Phys. B}

\def\de{\partial}

\newcommand{\mean}[1]{\langle #1 \rangle}

\begin{document}

\preprint{DESY 16-094}

\bibliographystyle{apsrev4-1}

\title{Probing axions with the neutrino signal from the next galactic supernova}

\author{Tobias~Fischer}
\affiliation{Institute for Theoretical Physics, University of Wroc\l{}aw, Pl. M. Borna 9, 50-204 Wroc\l{}aw, Poland}
\email{fischer@ift.uni.wroc.pl}

\author{Sovan Chakraborty}
\affiliation{Department of Physics, Indian Institute of Technology Guwahati, Assam 781039, India}
\affiliation{Max-Planck-Institut f{\"u}r Physik (Werner-Heisenberg-Institut), F{\"o}hringer Ring 6, 80805 M{\"u}nchen, Germany}

\author{Maurizio~Giannotti}
\affiliation{Physical Sciences, Barry University, 11300 NE 2nd Ave., Miami Shores, FL 33161, USA}

\author{Alessandro~Mirizzi} 
\affiliation{Dipartimento Interateneo di Fisica ``Michelangelo Merlin'', Via Amendola 173, 70126 Bari, Italy}
\affiliation{Istituto Nazionale di Fisica Nucleare - Sezione di Bari, Via Amendola 173, 70126 Bari, Italy}

\author{Alexandre~Payez}
\affiliation{Theory group, Deutsches Elektronen-Synchrotron DESY Notkestra\ss{}e 85, D-22607 Hamburg, Germany}

\author{Andreas~Ringwald}
\affiliation{Theory group, Deutsches Elektronen-Synchrotron DESY Notkestra\ss{}e 85, D-22607 Hamburg, Germany}

\date{\today}

\begin{abstract}
We study the impact of axion emission in simulations of massive star explosions, as an additional source of energy loss complementary to the standard neutrino emission. The inclusion of this channel shortens the cooling time of the nascent protoneutron star and hence the duration of the neutrino signal. We treat the axion-matter coupling strength as a free parameter to study its impact on the protoneutron star evolution as well as on the neutrino signal. We furthermore analyze the observability of the enhanced cooling in current and next-generation underground neutrino detectors, showing that values of the axion mass $m_a \gtrsim 8 \times 10^{-3}$~eV  can be probed. Therefore a galactic supernova neutrino observation would provide a valuable possibility to probe axion masses in a range within  reach of the planned helioscope experiment, the International Axion Observatory (IAXO).
\end{abstract}

\pacs{14.60.Pq,
97.60.Bw, 
14.80.Mz 
}

\maketitle

\section{Introduction} 

One of the most puzzling and long-standing problems in particle physics is related to the absence of an expected CP violation in the strong interactions: {\em The strong CP problem}. It is in this context and more precisely within the Peccei-Quinn mechanism~\cite{Peccei:1977hh,Peccei:1977ur,Weinberg:1977ma,Wilczek:1977pj} that axions, low-mass pseudoscalar particles with properties similar to those of neutral pions, have been introduced. Soon after the theoretical prediction of axions, it was moreover realized that such particles could be dark matter candidates in cosmology~\cite{Preskill:1983,Abbott:1983,Dine:1983}. Axions with masses on the order of $10^{-6}$~eV would behave as cold dark matter~\cite{Kawasaki:2013ae,Sikivie:2006ni,DiValentino:2014zna}, while for $m_a\agt60\times 10^{-3}$~eV they would attain thermal equilibrium at the QCD phase transition during the early universe expansion, or even later~\cite{Turner:1986tb,Masso:2002np}. In the latter case, axions would contribute to the cosmic radiation density and potentially to the cosmic hot-dark-matter density along with massive neutrinos~\cite{Archidiacono:2013cha}.

\begin{figure}[b!]
\subfigure[]{
\includegraphics[width=0.21\textwidth]{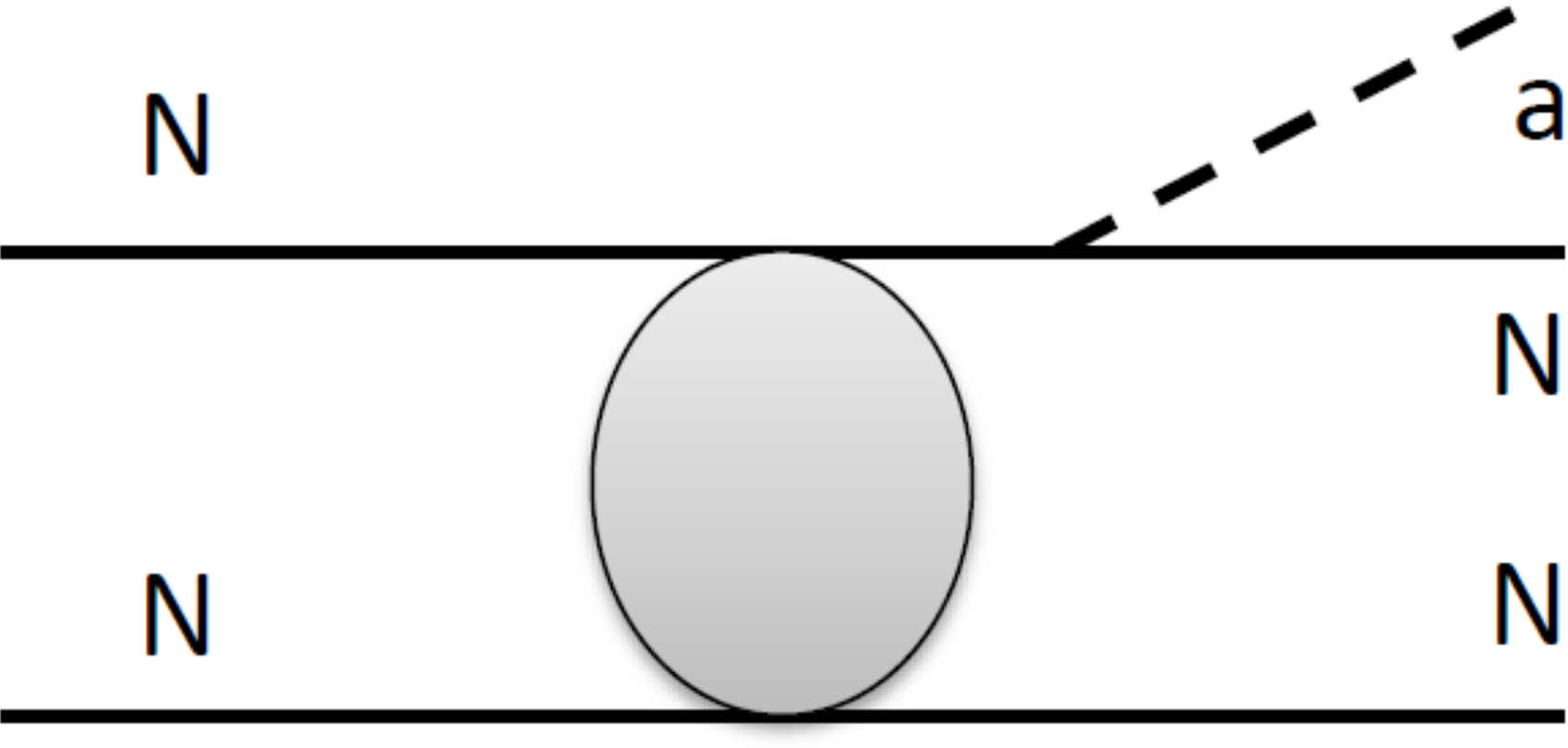}\label{fig:process_a}}
\hfill
\subfigure[]{
\includegraphics[width=0.21\textwidth]{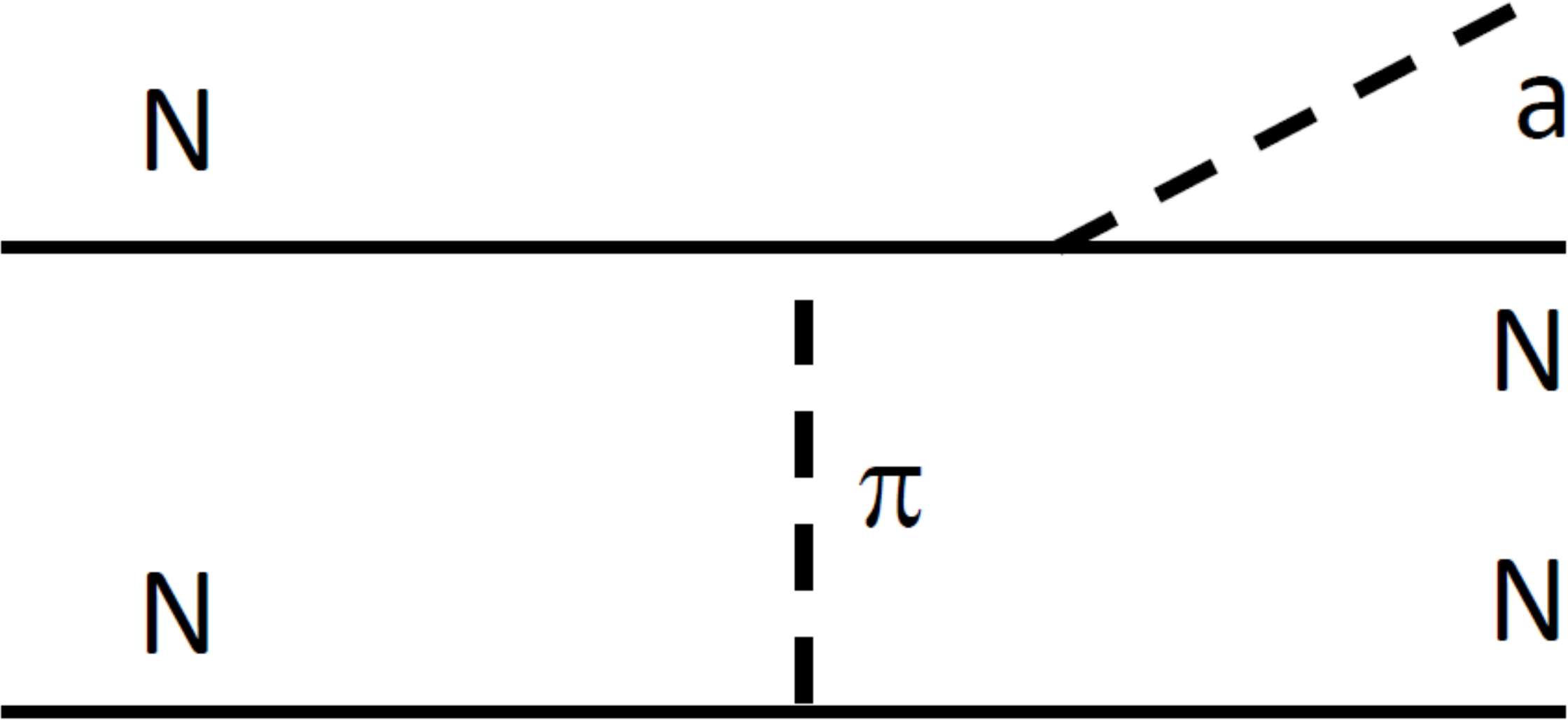}\label{fig:process_b}}
\caption{Example diagrams for the axion production through $N$--$N$ axion bremsstrahlung ($N=$ neutron or proton). The shaded region in \ref{fig:process_a} represents bulk nuclear interactions and \ref{fig:process_b} shows the contribution of the OPE approximation.}
\label{fig:process}
\end{figure}

The strongest bound on the axion mass comes from the observations of neutrinos originating from supernova (SN) SN1987A (cf. Refs.~\cite{Turner:1987by,Brinkmann:1988vi,Burrows:1988ah,Keil:1997,Raffelt:2006cw}). The relevant process of axion emission in a SN core is the nucleon--nucleon ($N$--$N$) axion bremsstrahlung, illustrated in Fig.~\ref{fig:process}, which involves the axion-nucleon coupling. Such an additional source of energy loss could potentially enhance the cooling, which in turn may reduce the associated neutrino flux. Our particular object of interest here is the protoneutron star (PNS), which forms when the imploding stellar core of a massive star reaches supersaturation density; the later ejection of the stellar mantle is subject to the explosion mechanism, and constitutes the SN problem. The PNS is initially hot and lepton rich, in which properties it differs from the final SN remnant: the neutron star. The deleptonization of the PNS is determined by the emission of neutrinos of all flavors, which decouple from matter at the neutrinospheres of last scattering, on a timescale of the order of 10--30~s. The observed duration of the neutrino burst from SN1987A was ${\mathcal O}$(10~s), in qualitative agreement with the expectations from ``standard'' SN models (for a recent review cf. Ref.~\cite{Janka:2012}). As a consequence, the upper bound on axion masses ranges between $5 \times 10^{-2}-6 \times 10^{-3}$~eV depending on the axion model~\cite{Keil:1997}. However, the sparse data sample of neutrino events from SN1987A and the currently still poor understanding of the nuclear medium at SN conditions suggest taking this limit as general guideline rather than a hard constraint.

Following this observation, we conduct for the first time consistent general relativistic neutrino radiation hydrodynamics simulations of the PNS deleptonization phase up to 40~s, with accurate three-flavor Boltzmann neutrino transport. We include $N$--$N$ axion bremsstrahlung at the level of the vacuum one-pion exchange (OPE) approximation taking, however, into account many-body effects. The latter suppress the rate of axion emission towards high densities~\cite{Raffelt:1991pw,Raffelt:1995,Janka:1996}. Similarly, the emission of neutrino--antineutrino pairs from $N$--$N$ bremsstrahlung can be calculated in the vacuum OPE approximation~\cite{Friman:1979}, with the addition of many-body effects~\cite{Hannestad:1997gc}. Beyond the OPE approach, medium modifications further suppress the rate with increasing density as shown recently using a chiral effective field theory approach~\cite{Bartl:2014,Bartl:2016} as well as based on the Fermi-liquid approach~\cite{Fischer:2016b}.

In this work we aim at studying the impact of the axion emission on the PNS evolution, as well as on the associated neutrino signal. Our results are in qualitative agreement with earlier studies~\cite{Keil:1997}. Current and future underground neutrino detectors guarantee a high statistics for the next galactic SN  event within the sensitivity range (see Ref.~\cite{Mirizzi:2015eza} for a recent review). Therefore, based on our new SN simulations, we calculate the neutrino events for the water-Cherenkov detector Super-Kamiokande and its future megatonne upgrade, as well as for the Cherenkov detector in the antarctic ice, Icecube. 

The plan of our work is as follows. In Sec.~II, we introduce our SN model AGILE-BOLTZTRAN and we discuss our reference simulation  without axions. In Sec.~III, we present the theoretical framework for the calculation of the axion emission rate, which is implemented in Sec.~IV into our SN simulations for which we discuss the evolution in comparison to the reference case. We explore the sensitivity of our results to the stellar model and to nucleon degeneracy. In Sec.~V, we study the impact of the axion emission on the observable neutrino signal in large underground detectors and show that values of $m_a \gtrsim  8 \times 10^{-3}$~eV  can be probed. The paper closes with the summary in Sec.~VI.

\section{Core-collapse SN simulations}

\subsection{Core-collapse SN model}
\label{SNmodel}

\begin{table}[htp!]
\centering
\caption{Neutrino reactions considered, including references.}
\begin{tabular}{ccc}
\hline
\hline
& Weak process & References \\
\hline
1 & $e^- + p \rightleftarrows n + \nu_e$ & \cite{Reddy:1998,Horowitz:2001xf} \\ 
2 & $e^+ + n \rightleftarrows p + \bar\nu_e$ & \cite{Reddy:1998,Horowitz:2001xf} \\
3 & $e^- + (A,Z) \rightleftarrows (A,Z-1) + \nu_e$ & \cite{Juodagalvis:2010} \\
4 & $\nu + N \rightleftarrows \nu' + N$ & \cite{Bruenn:1985en,Mezzacappa:1993gm,Horowitz:2001xf} \\
5 & $\nu + (A,Z) \rightleftarrows \nu' + (A,Z)$ & \cite{Bruenn:1985en,Mezzacappa:1993gm} \\
6 & $\nu + e^\pm \rightleftarrows \nu' + e^\pm$ & \cite{Bruenn:1985en},
\cite{Mezzacappa:1993gx} \\
7 & $e^- + e^+ \rightleftarrows  \nu + \bar{\nu}$ & \cite{Bruenn:1985en} \\
8 & $N + N \rightleftarrows  \nu + \bar{\nu} + N + N $ & \cite{Hannestad:1997gc} \\
9 & $\nu_e + \bar\nu_e \rightleftarrows  \nu_{\mu/\tau} + \bar\nu_{\mu/\tau}$ & \cite{Buras:2002wt,Fischer:2009} \\
10 & $(A,Z)^* \rightleftarrows (A,Z) + \nu + \bar\nu$ & \cite{Fuller:1991,Fischer:2013} \\
\hline
\end{tabular}
\\
$\nu=\{\nu_e,\bar{\nu}_e,\nu_{\mu/\tau},\bar{\nu}_{\mu/\tau}\}$ and $N=\{n,p\}$
\label{tab:nu-reactions}
\end{table}

In this study the spherically symmetric core-collapse supernova model AGILE-BOLTZTRAN is employed. It is based on general relativistic neutrino radiation hydrodynamics with angle-and energy-dependent three flavor Boltzmann neutrino transport ~\cite{Liebendoerfer:2001a,Liebendoerfer:2001b,Liebendoerfer:2002,Liebendoerfer:2004,Liebendoerfer:2005a}. Here we adopt the nuclear equation of state (EoS) from Ref.~\cite{Hempel:2009mc}, henceforth denoted as HS. Nuclei are treated within the modified nuclear statistical equilibrium approach for several 1000 nuclear species based on tabulated and partly calculated nuclear masses. The transition to homogeneous matter, with neutrons and protons only at densities in excess of normal nuclear matter density ($\rho_0$) and towards high temperatures around $T\simeq10-20$~MeV, is modeled intrinsically via a geometrical excluded volume approach based on the relativistic mean-field (RMF) framework. Here we select the RMF parametrization DD2 from Ref.~\cite{Typel:2005}; the final EoS is henceforth denoted as HS(DD2). In addition, lepton and photon contributions are calculated using the EoS from Ref.~\cite{Timmes:1999}.
 
The set of weak reactions considered is listed in Table~\ref{tab:nu-reactions}. For the weak processes with nucleons, both for charged-current absorption (reactions (1) and (2) in Table~\ref{tab:nu-reactions}) and for neutral-current scattering (reaction (4) in Table~\ref{tab:nu-reactions}), we employ here the elastic approximation~\cite{Bruenn:1985en}. Medium modifications for the charged current reactions~(1) and (2) in Table~\ref{tab:nu-reactions} are taken into account at the mean-field level. Therefore, the non-relativistic expressions of Ref~\cite{Reddy:1998}, Eq.\ (34), are modified in terms of mean-field potentials given by the nuclear EoS HS(DD2), neglecting medium dependent masses. Our medium modifications for the charged current reactions are introduced in Ref.~\cite{MartinezPinedo:2012}. They determine spectral differences between $\nu_e$ and $\bar\nu_e$~\cite{MartinezPinedo:2012,Roberts:2012,Horowitz:2012}, in particular for simulations of the PNS deleptonization. Moreover, the elastic (only momentum transfer) rate expressions for neutrino nucleon scattering of Ref.~\cite{Bruenn:1985en} are modified by the multiplicative neutrino energy-dependent factors of Ref.~\cite{Horowitz:2001xf}, which mimic modifications of the neutrino spectra due to inelastic contributions and weak magnetism corrections. Inelastic contributions are known to reduce both $\nu_e$ and $\bar\nu_e$ opacity, weak-magnetism corrections tend to generally increases differences between neutrinos and antineutrinos. Both effects have been commonly included in core-collapse SN simulations~\cite{Liebendoerfer:2005a,Buras:2005rp,Huedepohl:2010}. Beyond the mean-field effects, e.g. correlations are not considered here. They can be treated at the level of the random-phase approximation (RPA) as well as considering two-particle reactions, known as modified Urca processes \cite{Roberts:2012}. In the simulations, these effects result in small corrections of the neutrino fluxes and spectra during the late-time evolution of the deleptonization of the nascent PNS (see sec.~\ref{SNref}), in particular when neutrinos decouple at high densities.

Weak processes with heavy nuclei, i.e. electron captures, scattering and nuclear (de)excitation -- reactions (3), (5) and (10) in Table~\ref{tab:nu-reactions} -- are only important when nuclei are abundant. This is only the case during the core-collapse phase, when the temperature and the entropy per baryon are low. Once the entropy rises during the early post bounce evolution due to the presence of the bounce shock wave, material dissociates into bulk nuclear matter (neutrons, protons and light nuclei), and even into fully dissociated matter (neutrons and protons only). Consequently, during the post-bounce phase weak reactions with neutrons and protons are of relevance, including the PNS deleptonization after the explosion onset has been launched.

\subsection{Reference simulation -- evolution and neutrino signal}
\label{SNref}
Our SN simulations are launched from the 18.0~M$_\odot$ and 11.2~M$_\odot$ pre-collapse progenitors of Ref.~\cite{Woosley:2002zz}, henceforth denoted as s18 and s11.2 respectively. Both stellar models were evolved consistently through all SN phases. The neutrino signal as well as the SN shock dynamics and the neutrinospheres (for all flavors) are illustrated in Figs.~\ref{fig:neutrinos_ref} and \ref{fig:shock_spheres} for s18. The evolution is in qualitative agreement with s11.2 for which results are partly discussed in Ref.~\cite{MartinezPinedo:2014}.

\subsubsection{Stellar core collapse}
The stellar core collapse is triggered by the loss of pressure from the degenerate electron gas, as the electrons are captured on protons bound in heavy nuclei. It is therefore essential to include electron-capture rates based on detailed microscopic nuclear models, as discussed in details in the literature~\cite{Hix:2003,Langanke:2003ii}, for the deleptonization during core collapse. These rates determine the lepton fraction ($Y_L$) of the stellar core, which equals the electron fraction $Y_e$ until neutrinos become trapped, after which $Y_e<Y_L$. The further evolution of the electron fraction $Y_e$ beyond neutrino trapping is mainly determined by the nuclear symmetry energy~\cite{Fischer:2014}. 

Since nuclear electron captures produce only $\nu_e$, the $\nu_e$ luminosity and the average energy rise during core collapse (see Fig.~\ref{fig:neutrinos_ref}). An additional source of neutrinos was proposed in Ref.~\cite{Fuller:1991}, via the de-excitation of excited nuclear states and the emission of neutrino pairs; reaction~(10) in Table~\ref{tab:nu-reactions}. It is based on the presence of excited nuclear states due to the temperatures reached during collapse on the order of few MeV. This has been recently explored in SN simulations~\cite{Fischer:2013}. However, the (de)excitation rates are much smaller than those of electron captures and consequently the observed luminosities are small compared to those of $\nu_e$. Hence the influence on the core-collapse evolution is  negligible.

\subsubsection{Core bounce and post bounce evolution}

In the final phase of the stellar core collapse, the density exceeds $\rho_0$ when the strong short-range repulsive nuclear force counterbalances gravity. This halts the collapse with the formation of a strong hydrodynamics shock wave (green solid line in Fig.~\ref{fig:shock_spheres}). It propagates rapidly to large radii on the order of 50--180~km, with the PNS enclosed. The latter is initially very dilute, being hot and lepton rich. In these two latter properties the PNS differs from the final SN remnant, i.e.\@ a neutron star.

The deleptonization burst in the upper-left panel of Fig.~\ref{fig:neutrinos_ref} is associated with the bounce shock propagation across the $\nu_e$-sphere of last inelastic scattering (black solid line in Fig.~\ref{fig:shock_spheres}) where a large number of electron captures on free protons releases this $\nu_e$-burst. The shock stalling due to this energy loss, accompanied by the dissociation of infalling heavy nuclei from the still gravitationally unstable layers above the stellar core, results in the post-bounce mass accretion phase. Thereby a thick low-density layer of accumulated material develops at the PNS surface, in which $\nu_e$ and $\bar\nu_e$ decouple. The high magnitude of their luminosities  -- on the order of $10^{52}$~erg~s$^{-1}$ -- is determined by the mass accretion rate. On the other hand, the heavy-lepton flavor neutrinos decouple at generally higher densities due to the absence of charged-current absorption reactions. The different coupling strength to matter is reflected in the hierarchy of their average energies during the accretion phase, $\langle E_{\bar\nu_{\mu/\tau}} \rangle>\langle E_{\nu_{\mu/\tau}} \rangle>\langle E_{\bar\nu_{e}} \rangle>\langle E_{\nu_{e}} \rangle$ (cf. Ref.~\cite{Raffelt:2001} and the bottom panels in Fig.~\ref{fig:neutrinos_ref}).

\subsubsection{Shock revival and explosion onset}

The evolution of the $\nu_e$ and $\bar\nu_e$ luminosities during the mass accretion phase reflects oscillations of the bounce shock and hence of the mass accretion rate at the PNS surface~\cite{Miller:1993,Buras:2005rp}. These are in part associated with the enhanced neutrino heating treatment which we apply here in order to trigger the SN explosion onset, i.e. the expansion of the bounce shock to increasingly larger radii. Thereby the heating rates for reactions~(1) and (2) of Table~\ref{tab:nu-reactions} are increased inside the heating region. This method has been employed previously~\cite{Fischer:2009af,MartinezPinedo:2012}; it compares well with other artificially neutrino-driven explosion methods~\cite{Ugliano:2012,Perego:2015}, which is necessary because in spherically symmetric simulations neutrino-driven explosions cannot be obtained except for very light progenitor stars~\cite{Kitaura:2006,Melson:2015}. Here, it results in the slow but continuous expansion of the bounce shock to increasingly larger radii (see Fig.~\ref{fig:shock_spheres}), with the onset of the explosion around $t=0.25$~s post bounce for the s18 and at about $t=0.15$~s for s11.2. The explosion shock reaches radii around 1000~km at about $t=0.5$~s post bounce. Once the explosion proceeds, we switch back to the standard rates.

\subsubsection{PNS deleptonization}

\begin{figure}[htp!]
\includegraphics[width=0.5\textwidth]{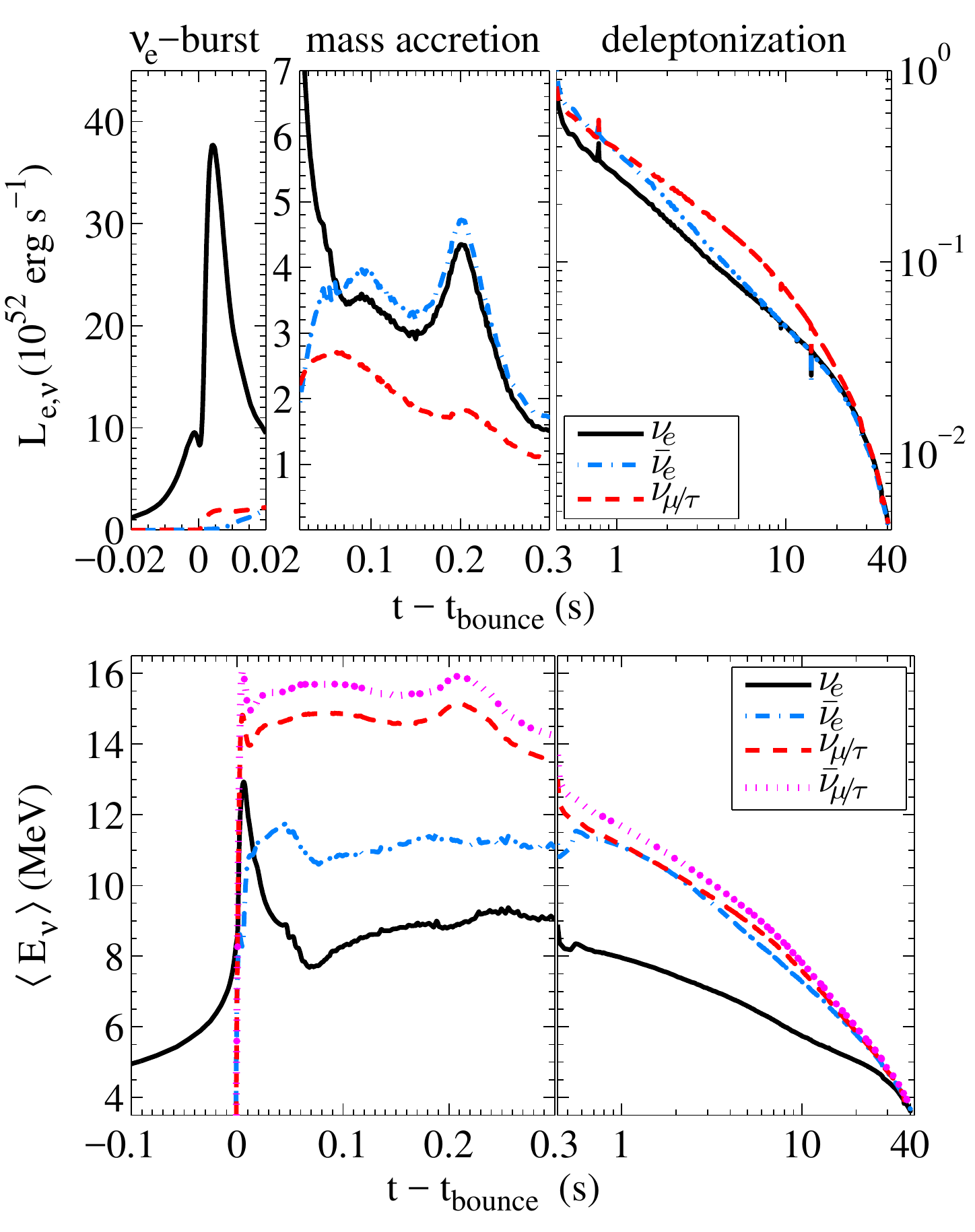}
\caption{(color online) Evolution of neutrino energy luminosities (top panel) and average energies (bottom panel) for the reference simulation of s18.}
\label{fig:neutrinos_ref}
\end{figure}

\begin{figure}[htp!]
\includegraphics[width=0.5\textwidth]{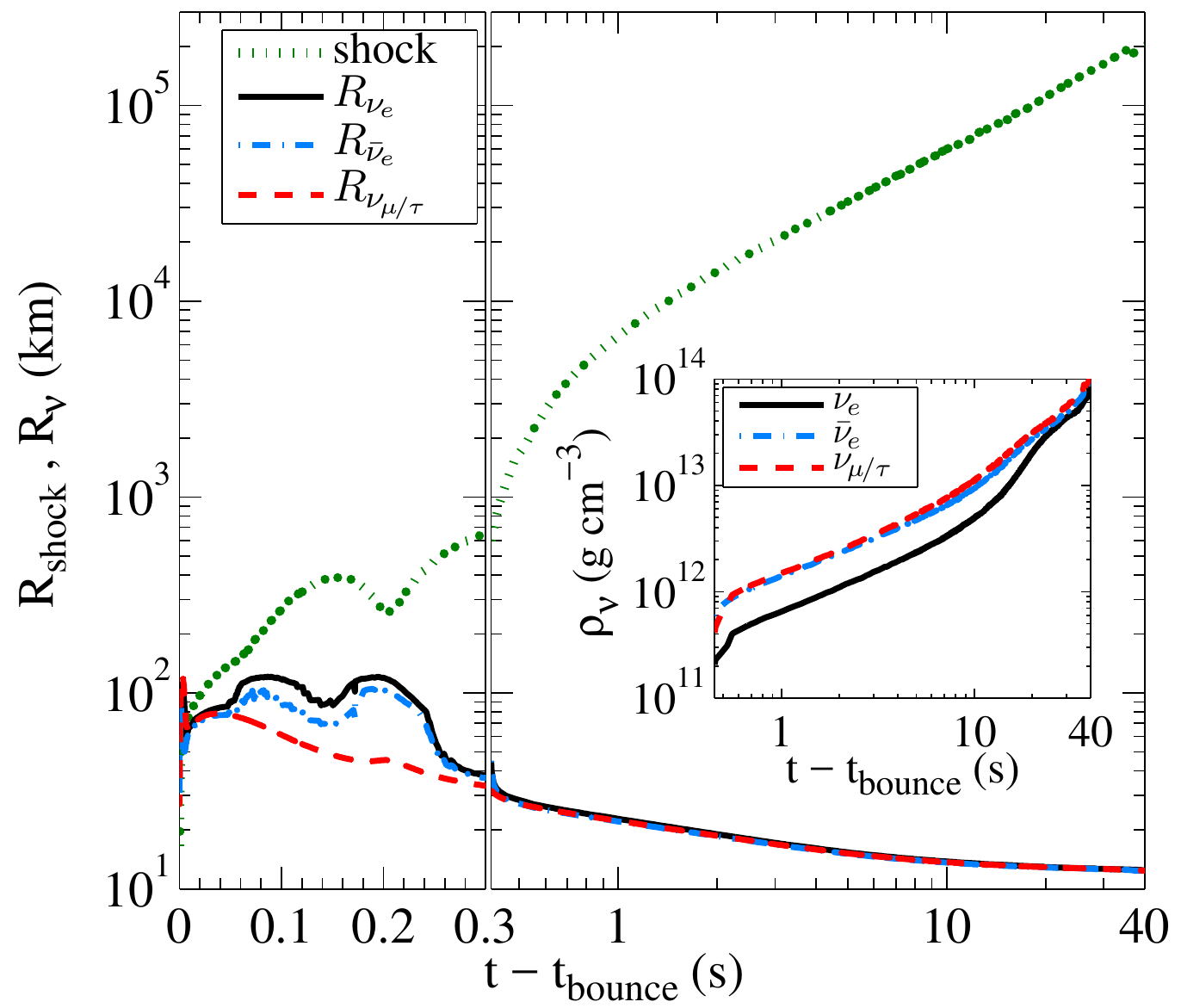}
\caption{(color online) Evolution of shock radius and neutrinospheres for the reference simulation of s18. The inlay shows the density at the corresponding neutrinospheres.}
\label{fig:shock_spheres}
\end{figure}

In spherically symmetric models, with the shock revival mass accretion vanishes completely at the PNS surface, and the neutrino fluxes turn rapidly from accretion dominated towards diffusion (see Fig.~\ref{fig:neutrinos_ref}). It has been demonstrated in multidimensional simulations that there is an extended transition period during which the presence aspherical flows enhances the luminosities above the diffusion limit~\cite{Mueller:2014}. However, the long-term evolution of the nascent PNS cannot be studied in multi-dimensional simulations. Here, the neutrino fluxes drop by one order of magnitude during the first second after the explosion onset and they become increasingly similar towards later times of the PNS deleptonization. Fig.~\ref{fig:shock_spheres} illustrates the evolution of the corresponding neutrinospheres. Moreover the initial neutrino energy hierarchy is broken, with $\langle E_{\nu_{\mu/\tau}} \rangle\simeq\langle E_{\bar\nu_{e}} \rangle>\langle E_{\nu_{e}} \rangle$. This is due to the reduced importance of charged current absorption reactions for $\bar\nu_e$ during the PNS deleptonization phase. Instead, the opacity of heavy lepton flavor neutrinos and $\bar\nu_e$ are determined by the same set of weak processes, dominated by elastic scattering on neutrons. Hence their spectra become increasingly similar, unlike $\nu_e$ in which the opacity is continuously dominated by charged-current absorption on neutrons. This property is a general feature of the PNS deleptonization and has been recognized and analyzed~\cite{Fischer:2012a}. 


Note that the PNS deleptonization phase is mildly independent from the details of the SN explosion mechanism and the ejection of the stellar mantle. On the other hand, PNS convection as well correlations of the nuclear medium that influence the weak processes, both of which modify the PNS deleptonization~\cite{Roberts:2012f,Roberts:2012,Mirizzi:2015eza}, are not included here.

\section{Axion production} 

To study the effect of the additional axion cooling on the neutrino signal, we have to evaluate the axion production rate in a newly born SN environment and the energy carried away by those axions. In general, axions can be produced through electromagnetic processes, namely the Primakoff production~\cite{Payez:2014xsa} and the axion-electron bremsstrahlung~\cite{Ellis:1987pk}, and through nuclear processes, specifically the $N$--$N$ axion bremsstrahlung:
\begin{eqnarray}
N_1 + N_2 \longrightarrow N_3 + N_4 + a\,,
\label{eq:brems}
\end{eqnarray}
shown in Fig.~\ref{fig:process}, where $N_i $ are nucleons (protons or neutrons) and $ a $ is the axion field. In this work we will focus on the process \eqref{eq:brems}, which is the dominant axion production mechanism in the hot and dense environment characterizing the core of a newly born SN~\cite{Turner:1987by}.

The process \eqref{eq:brems} is induced by the axion-nucleon interaction described by the following Lagrangian term,
\begin{eqnarray}\label{eq:axion_N_coupling}
\mathcal{L}_{a N}=\sum_{i=p,n} \frac{g_{a i}}{2m_N}\,\overline N_i \gamma_\mu\gamma_5 N_i\de^\mu a,
\end{eqnarray}
with axion-nucleon couplings defined as follows,
\begin{equation}
g_{ai} =C_i \frac{m_N}{f_a} = 1.56 \times 10^{-7} \left(\frac{m_a}{\textrm{eV}} \right)C_i~,
\label{eq:coupl}
\end{equation}
where $f_a$ is the Peccei-Quinn energy scale, $C_i$ are model dependent constants and $ m_N$ is the nucleon mass (we assume $ m_n\simeq m_p $). In the right-hand side of the previous equation we used the relation between $f_a$ and the axion mass $m_a$,
\begin{equation}
m_a = 0.60~{\rm eV} \left(\frac{10^7 {\rm GeV}}{f_a}\right)~.
\label{eq:mass}
\end{equation}

In the case of the hadronic axion model or KSVZ (Kim-Shifman-Vainshtein-Zakharov)~\cite{Kim:1979if,Shifman:1979if} and DFSZ (Dine-Fischler-Srednicki-Zhitnitsky)~\cite{Dine:1981rt,Zhitnitsky:1980tq}, the constants $C_i$ have recently been computed with remarkable accuracy combining next-to-leading order chiral perturbation theory with Lattice QCD~\cite{diCortona:2015ldu}:
\begin{eqnarray}
\label{Eq:lattice_couplings}
	&& C_p^{\rm (KSVZ)}=-0.47\pm 0.03\,, \nonumber \\
	&& C_n^{\rm (KSVZ)}=-0.02\pm 0.03\,. \nonumber \\
	&& C_p^{\rm (DFSZ)}=\left(-0.182\pm 0.025\right)-0.435\cos^2\beta\,, \nonumber \\
	&& C_n^{\rm (DFSZ)}=\left(-0.16\pm 0.025\right)+0.414\cos^2\beta\,,
\end{eqnarray}
where $ \tan\beta $ is the ratio between the Higgs doublets in the DFSZ model.

Interestingly, from the above expressions we notice that neither model allows the proton coupling to vanish within the errors, while the coupling to neutrons is compatible with zero in the KSVZ model and also in the DFSZ model if $\cos^2 \beta\sim 0.4$. Therefore, as a benchmark for our analysis, we will consider only interactions with protons in our simulations though we will provide all the necessary relations for the most general case.  

The nuclear axion bremsstrahlung rate is highly uncertain, mostly due to the lack of understanding of the nuclear interactions; approximations are commonly applied based on vacuum physics. A fundamental consequence of the nucleon-axion interaction (\ref{eq:axion_N_coupling}) is that the nucleon spins flip in collisions and so spin-conserving interactions do not contribute to the axion bremsstrahlung production (cf. Ref.~\cite{Raffelt:2006cw}). Any description of the nuclear interaction in relation to the axion emission process has to account for these results.

Progress in this direction has been possible with the introduction of the spin-density structure function formalism~\cite{Iwamoto:1982zp,Raffelt:1991pw}. The functions describe the correlations of the spin density operators (see, e.g., \cite{Iwamoto:1982zp,Raffelt:1991pw,Janka:1996}). They contain the nuclear part of the matrix element squared and include all the expected many-body effects. However, practically the matrix elements can only be calculated in specific frameworks, the most widely used being the OPE potential, which describes the two nucleon interaction with the exchange of a pion (see the right panel of Fig.~\ref{fig:process}).

This interaction is described by the following effective vertex,
\begin{eqnarray}
\mathcal{L}_{\pi N}=(2 m_N f_{ij}/m_{\pi})\overline N_i \gamma_5 N_j \pi~, 
\end{eqnarray}
where $f_{ij}\sim 1$ is a phenomenological constant ($i,j=n,p$) which depends on whether the mediator is a $\pi^0$ or a $\pi^\pm$, being $f_{np}=\sqrt 2 f_{nn}=-\sqrt 2f_{pp}$, as required by the isospin invariance. In general, the nucleon-pion interaction has the derivative form $(f_{ij}/m_{\pi})\overline N_i \gamma_\mu\gamma_5 N_j\de^\mu \pi$, typical of the (pseudo-) Goldstone modes, just as the axion. However, this interaction can be made pseudoscalar (as in the main text), after an opportune chiral rotation of the nucleon fields. Yet, this operation cannot be performed for both pion and axion fields simultaneously (for more details cf. Ref.~\cite{Brinkmann:1988vi} and references therein).

Though the OPE approximation is a good starting point for the description of the axion bremsstrahlung, it does oversimplify the nuclear dynamics and overestimates the emission rate~\cite{Hanhart:2000ae}. Here, we refer the reader to the extended literature~\cite{Raffelt:1991pw,Raffelt:1993ix,Janka:1995ir,Keil:1997,Hanhart:2000ae,Raffelt:2006cw} for the peculiarities of the axion emission rate in a nuclear medium. In general, a reliable framework to extract the details of the axion emission rate from a SN core is still missing and we rely on approximate descriptions in order to better compare with previous works. In the present study, we follow the procedure described in Ref.~\cite{Keil:1997} and implement the derived rate in our numerical SN model. The procedure assumes a modified OPE potential to account for many-body effects and the subsequently reduced axion production rate with increasing density. However, we remark that even this approximation is subject to essentially unquantifiable uncertainties. 

In Ref.~\cite{Keil:1997}, the volume axion rate is calculated as follows,
\begin{eqnarray}
Q_a(T,\rho,\mu_n,\mu_p)=Q_a^{(1)}{\rm min} \left[ 1,\frac{\Gamma_\sigma^{\rm max}}{\Gamma_\sigma^{(1)}}  \right]~,
\label{eq:Qa_gamma}
\end{eqnarray}
as a function of temperature $T$, density $\rho$ and the nucleon chemical potentials $\mu_{n,p}$, where the overall magnitude is determined via the following relation,
\begin{widetext}
\begin{eqnarray}
Q_a^{(1)} &=& \int \frac{d^3p_a}{2E_a(2\pi)^3} \prod_{i=1,4} \frac{d^3p_i}{2E_i(2\pi)^3}~E_a f_1 f_2 (1-f_3) (1-f_4)
\sum_\text{spins} \vert \mathcal{M} \vert ^2
\delta^4(p_1+p_2-p_3-p_4-p_a) \nonumber 
\\
&\simeq& 64\left(\frac{f}{m_\pi}\right)^4
\left(\frac{m_N^{5/2}\,T^{13/2}}{\rho}\right)
\left\{
\left(1-\frac{\xi}{3}\right)\,g_{\rm an}^2\,I(y_n,y_n) +
\left(1-\frac{\xi}{3}\right)\,g_{\rm an}^2\,I(y_p,y_p)
\right.
\nonumber
\\
&&
\left.
+ \frac{4(15-2\xi)}{9}\left(\frac{g_{\rm an}^2+g_{\rm ap}^2}{2}\right)\,I(y_n,y_p) +
\frac{4(6-4\xi)}{9}\left(\frac{g_{\rm an}+g_{\rm ap}}{2}\right)^2\,I(y_n,y_p)
\right\}~.
\label{eq:Qa}
\end{eqnarray}
\end{widetext}
In Eq.~\eqref{eq:Qa_gamma}, the term
\begin{eqnarray}
\Gamma_\sigma^{(1)}=10 \, {\rm MeV}\,  \left( \frac{m_N}{938\,{\rm MeV}}\right)^2 \rho_{14}\,T_{\rm MeV}^{1/2}\,,
\end{eqnarray}
describes the lowest-order effective spin fluctuation rate~\cite{Keil:1997}, with $ \rho_{14}=\rho/10^{14}~\text{g cm}^{-3} $ and $ T_{\rm MeV}=T/ \text{MeV}$. Finally, following Ref.~\cite{Keil:1997}, we select an average value of $\Gamma_\sigma^{\rm max}=60~{\rm MeV}$ which accounts for the saturation of $\Gamma_\sigma$. 

The matrix elements $|\mathcal{M}|^2$ in Eq.~\eqref{eq:Qa} are calculated in the vacuum OPE framework. The fitting functions $I(y_1,y_2)$ are given in Eq.~(13) of Ref.~\cite{Keil:1997}, with nucleon degeneracy $y_i=\mu_i^0/T$ and nucleon chemical potentials $\mu_i^0 = \mu_i-m_i$, where we are assuming bare nucleons masses. Finally, the degeneracy parameter $\xi$ in Eq.~\eqref{eq:Qa} attains the values $\xi=0$ and $\xi=1$ in the limits of fully degenerate and non-degenerate nucleons  respectively; it is defined as follows~\cite{Brinkmann:1988vi,Keil:1997},
\begin{eqnarray}
\xi= \langle |\hat{\textbf{ k}}\cdot \hat{\textbf{l}}|^2 \rangle~,
\end{eqnarray}
where $\textbf{k}=\textbf{p}_2-\textbf{p}_4$ and $\textbf{l}=\textbf{p}_2-\textbf{p}_3$ indicate the momentum transfers ($ \textbf{p}_i $ is the nucleon momentum) in the direct and exchange scattering diagrams. The  effective spin fluctuation rate in Eq.~(\ref{eq:Qa_gamma}), with the saturation term, suppresses the OPE production rate at high densities, correcting the ill behaving OPE approximation at short distance. Moreover, the derivation of Eq.~\eqref{eq:Qa} is based on the assumption of freely streaming axions once they are produced, i.e. the possibility of any reabsorption and/or scattering of axions after they are produced is ignored. This is a condition easily satisfied for axions with coupling to nucleons in the range we are interested in, $g_{ai}\simeq 10^{-10}$, as in this case the typical axion mean free path is several orders of magnitude larger than the SN radius~\cite{Burrows:1990pk}. 

Possible effects of the medium induced modification of the nuclear mass have also been neglected. Axions are produced mostly from regions with high temperature while high density, though relevant, has a minor impact on the production rate. Numerical estimates show that most of the axion emission happens in the first couple of seconds during the PNS deleptonization, in a narrow region at the PNS center, where the density is never high enough to induce a modification of the nuclear mass by more than 30\%.

\begin{table*}[htp]
\centering
\caption{Data from PNS deleptonization}
\begin{tabular}{ccccccccccc}
\hline
\hline
Progenitor & $M_{\rm B}^{\rm PNS}$ & $M_{\rm G}^{\rm PNS}$ &  $g_{\rm ap}$ & $\xi$ & $\text{max}\left(L_{e,a}\right)$ & $t(\text{max}\left(L_{{\rm e},a}\right))$ & $L_{{\rm e},\nu}(10~\rm s)$  & $T_c(10~\rm s)$ & $E_\nu^{\rm tot}\vert_{t=20~\rm s}$ & $E_a^{\rm tot}\vert_{t=20~\rm s}$ \\
& $[$M$_\odot]$ & $[$M$_\odot]$ & $[10^{-10}]$ &   & $[10^{51}$~erg~s$^{-1}]$ & $[$s$]$ & $[10^{51}$~erg~s$^{-1}]$  & $[$MeV$]$ & $[10^{53}$~erg$]$ & $[10^{53}$~erg$]$ \\
\hline
s18 & 1.62 & 1.46 & 0 & - & 0 & - & 3.8 & 38.6 & 2.0 & - \\
 &  &   & 9 & 1 & 30.0 & 1.6 & 1.6 & 16.6 & 1.16 & 0.95 \\
 &  &   & 9 & 0 & 37.2 & 1.5 & 1.4 & 15.8 & 1.17 & 0.95 \\
 &  &   & 6 & 0 & 20.0 & 1.8 & 1.9 & 18.6 & 1.27 & 0.78  \\
\hline
s11.2 & 1.29 & 1.19 & 0 & - & 0 & - & 2.6 & 32.4 & 1.25 & - \\
 &  &   & 9 & 1 & 7.4 & 2.3 & 1.5 & 19.8 & 1.06 & 0.32 \\
\hline
\end{tabular}
\label{tab:data}
\end{table*}

\bigskip

\section{PNS evolution -- shortened deleptonization with axions}

In order to study the role of axions we implement the process~\eqref{eq:brems} into our SN model. Due to the generally low axion-nucleon coupling we assume that the emitted axions are freely streaming~\cite{Burrows:1988ah}, i.e.\@ no axion transport is required. We treat axions as a separate particle species in addition to baryons, leptons and photons. Hence axions cannot contribute to the equation of state, e.g, to energy density, entropy and pressure; however, they contribute to the cooling via the associated energy losses. The axion luminosity is calculated by integrating the local energy-loss rate Eq.~\eqref{eq:Qa_gamma},
\begin{equation}
L_{{\rm e},a} = \int_0^M dm\,Q_a(T,\rho,\mu_n,\mu_p)~,
\label{eq:La}
\end{equation}
over the enclosed baryon mass $m$ from the core towards the surface $M$. The associated losses are then treated as an additional sink term in the equation of energy conservation. It is evident that this expression depends only on the choice of ($g_{\rm ap}$, $\xi$) and $\Gamma_\sigma^{\rm max}$, besides the local conditions ($T,\rho,Y_e$). The dimensional analysis of \eqref{eq:Qa} gives a rough estimate of the local energy loss rate from axion production. Assuming $\Gamma_\sigma^{\rm max}/\Gamma_\sigma^{(1)}>1$ we estimate the total energy loss from axion emission,
\begin{widetext}
\begin{eqnarray}
L_{{\rm e},a} \sim 2.6\times 10^{51}
\,
\left(\frac{2\times10^{14} ~\rm g~cm^{-3}}{\rho}\right)
\left(\frac{T}{10~{\rm MeV}}\right)^{13/2} 
\left(\frac{g_{\rm ap}}{10^{-10}}\right)^2
\left(\frac{Y_p}{0.1}\right)^2
\left(\frac{M}{0.5~{\rm M}_\odot}\right)
\,
{\rm erg}~{\rm s}^{-1}~,
\label{eq:La_dim}
\end{eqnarray}
\end{widetext}
where we assumed $m_\pi=135$~MeV and $m_N=938$~MeV for the pion mass and nucleon mass respectively. Furthermore, we assumed $g_{\rm an}=0$ and the relation $I(y_p,y_p)\propto (Y_p)^2$ valid for abundance of targets with $Y_p=Y_e$. 
Eq.~\eqref{eq:La_dim} can be estimated using the average temperature and density. Notice, however, that Eq.~\eqref{eq:La_dim} ignores the axion feedback on the temperature, which is quite relevant for the axion couplings we are considering here, as clear from Fig.~\ref{fig:axion}. It should therefore be taken only as an estimate of the axion luminosity and of its dependence on the relevant physical quantities.

Due to the restriction to spherical symmetry, quantitative estimates about the role of axions in the supernova explosion dynamics, i.e. during the accretion phase prior to the revival of the stalled bounce shock, are not meaningful. The total energy released in neutrinos of all flavors during the accretion phase in our simulations is $E_\nu^{\rm tot}=0.7(+0.01)\times 10^{53}$~erg for s18 and $E_\nu^{\rm tot}=0.3(+0.008)\times 10^{53}$~erg for s11.2. Values in parentheses refer to the energy released during stellar collapse including the $\nu_e$-deleptonization burst between 5--20~ms post bounce. For comparison, we list the total energy emitted in neutrinos during the PNS deleptonization in Table~\ref{tab:data} -- the lines with $g_{\rm ap}=0$ correspond to the reference simulations. The PNS deleptonization phase corresponds to the period of the SN when most of the trapped neutrinos are being released. In the present study we are interested in the impact of the additional source of energy loss from axion emission on the structure and evolution of the PNS during the deleptonization, i.e.\@ after the explosion onset. 

\subsection{Comparison with the reference simulation}
In accordance with previous studies we neglect the neutron channel $g_{\rm an}=0$. It leaves $g_{\rm ap}$ as free parameter such that $g_{\rm an}/g_{\rm ap}$=0. In addition we assume non-degenerate protons ($\xi=1$) and as representative value of the axion-proton coupling strength we select $g_{\rm ap}=9\times 10^{-10}$ (comparable with the SN1987A bound) with a the saturation of $\Gamma_\sigma^{\rm max}=60$~MeV. This value of the $g_{\rm ap}$ corresponds to $m_a \simeq 3 \times 10^{-2}$~eV or $f_a \simeq  4.8 \times 10^{8}$~GeV [see Eqs.~\eqref{eq:coupl}--\eqref{eq:mass}].

\begin{figure}[b!]
\includegraphics[width=0.5\textwidth]{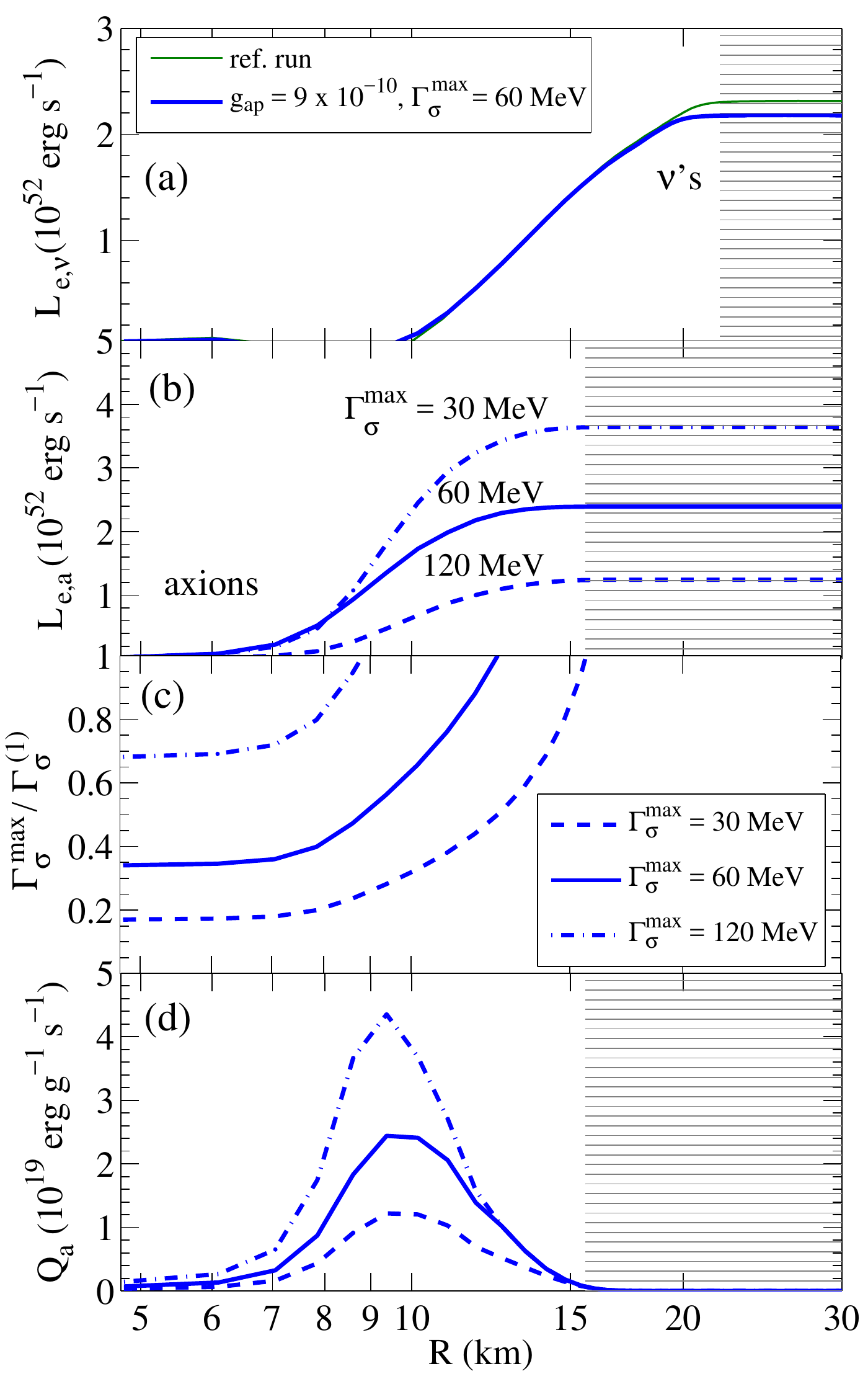}
\caption{(color online) Radial profiles of selected quantities at 1~s post bounce (see text for definitions).}
\label{fig:axion}
\end{figure}
\begin{figure*}[htp!]
\includegraphics[width=\textwidth]{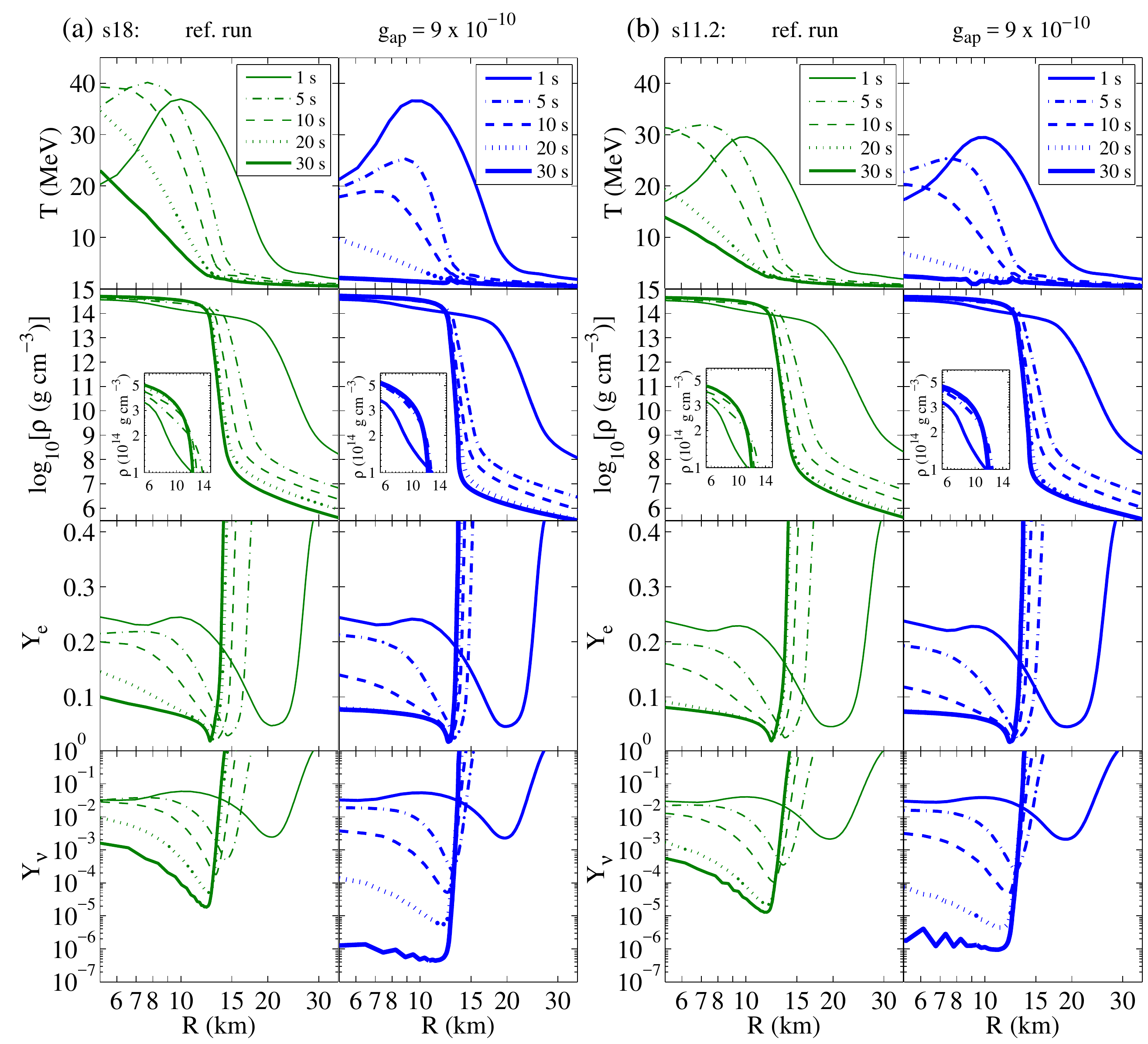}
\caption{(color online) Evolution of selected quantities -- from top to bottom: temperature, density, electron and neutrino abundances -- during the PNS deleptonization, comparing our reference simulations (left panels) with the simulation including axions (right panels) with ($g_{\rm ap}=9\times 10^{-10}, \xi=1$).}
\label{fig:hydro}
\end{figure*}

The SN simulations discussed below are launched with this parameter setup, unless stated otherwise. In Figure~\ref{fig:axion} we illustrate radial profiles of the total neutrino and axion luminosities in graphs~(a) and (b) as well as the local axion emission rate $Q_a$ in graph~(d), corresponding to conditions obtained at about 1~s post bounce during the early PNS deleptonization. The steeply rising neutrino luminosity corresponds to the region of neutrino decoupling (where all weak process have ceased), outside of which all the neutrino luminosity remains constant (see region marked by the gray-shaded area in graph (a)). The same holds for the axion emission, i.e.\@ the axion luminosity rises in the region where $Q_a>0$, and as $Q_a\to 0$ the axion luminosity remains constant, with no more axion production (marked by the gray-shaded area in graphs~(b) and (d)). Here it becomes evident that, unlike neutrinos, axions are emitted mainly from the PNS interior. It corresponds to the region with high densities and in particular with the highest temperatures. Note the high power of $T$ in Eq.~\eqref{eq:Qa} which explains this strong temperature dependence of the axion emission rate. With increasing distance from the center, the matter density, and consequently also the density of protons, reduces; it drops rapidly below $\rho\sim 10^{14}$~g~cm$^{-3}$ between $R=10$--$15$~km as illustrated at the example of radial profiles of selected quantities in Fig.~\ref{fig:hydro} graphs~(a) and (b). Consequently the axion emission rate drops rapidly to vanishing values with distance from the center [see Fig.~\ref{fig:axion})(d)].

\begin{figure}[htp!]
\subfigure[~Energy and number luminosities]{
\includegraphics[width=0.48\textwidth]{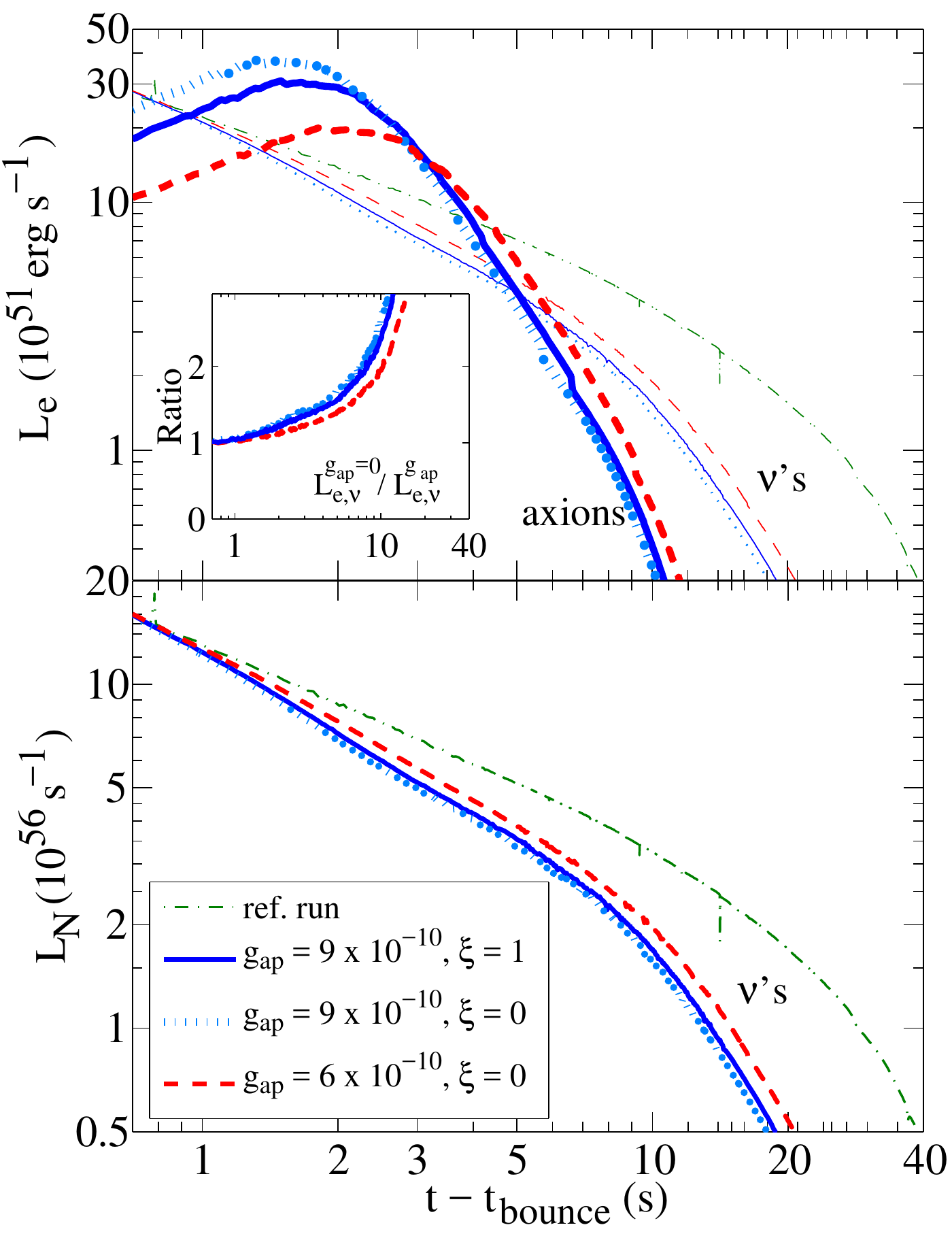}\label{fig:h18a_lumin_gap}}
\\
\subfigure[~Average neutrino energies]{
\includegraphics[width=0.48\textwidth]{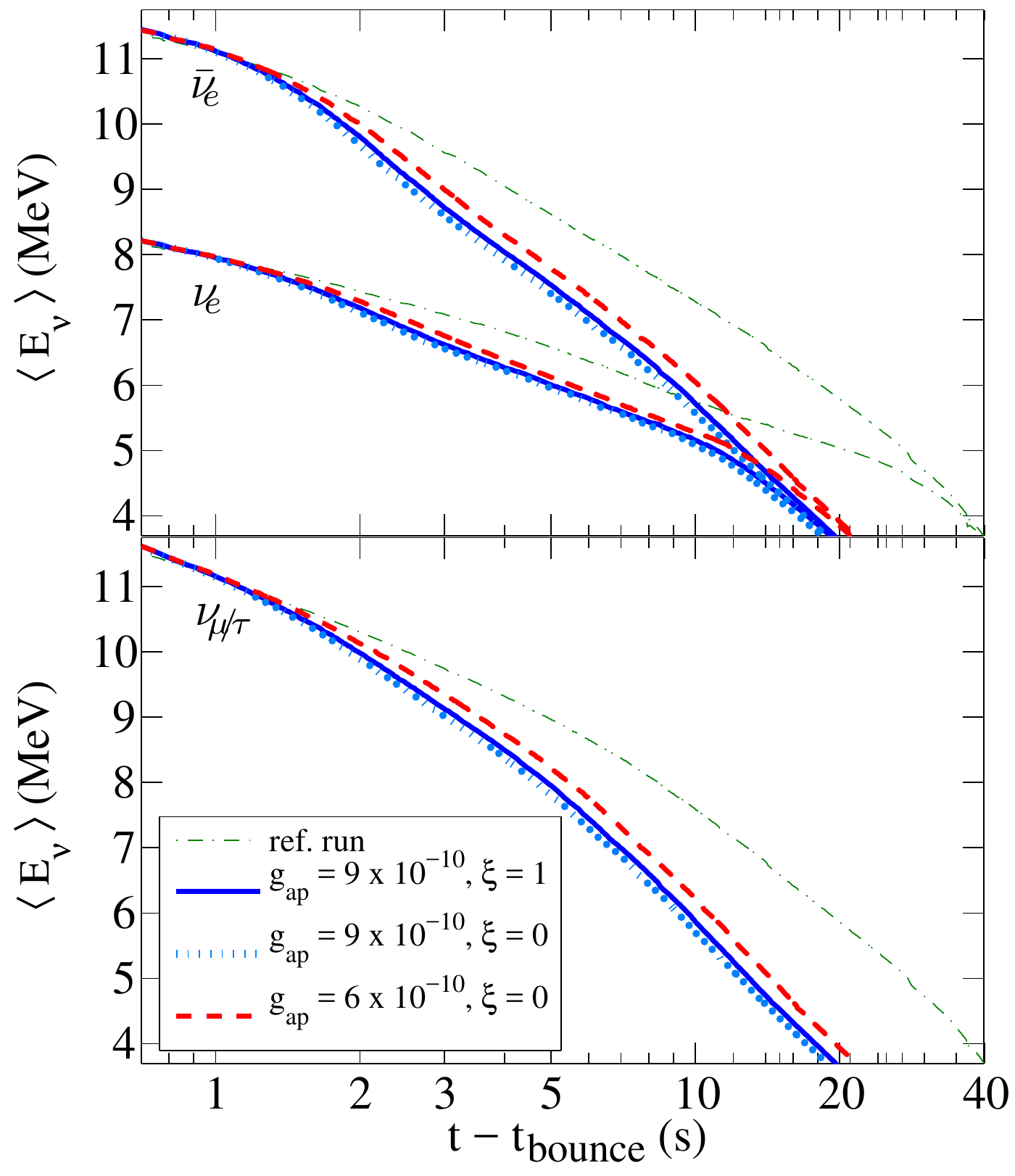}\label{fig:h18a_rms_gap}}
\caption{(color online) Neutrino signal for the simulations based on s18, comparing reference run (green	 dash-dotted lines) and simulations with axions for different values of ($g_{\rm ap},\xi$). The inlay in the top panel shows the ratios of $L_{\rm e,\nu}$ for the simulations with axions relative to the reference run with $g_{\rm ap} = 0$.}
\label{fig:neutrinos_s18}
\end{figure}
\begin{figure}[htp!]
\subfigure[~Energy and number luminosities]{
\includegraphics[width=0.48\textwidth]{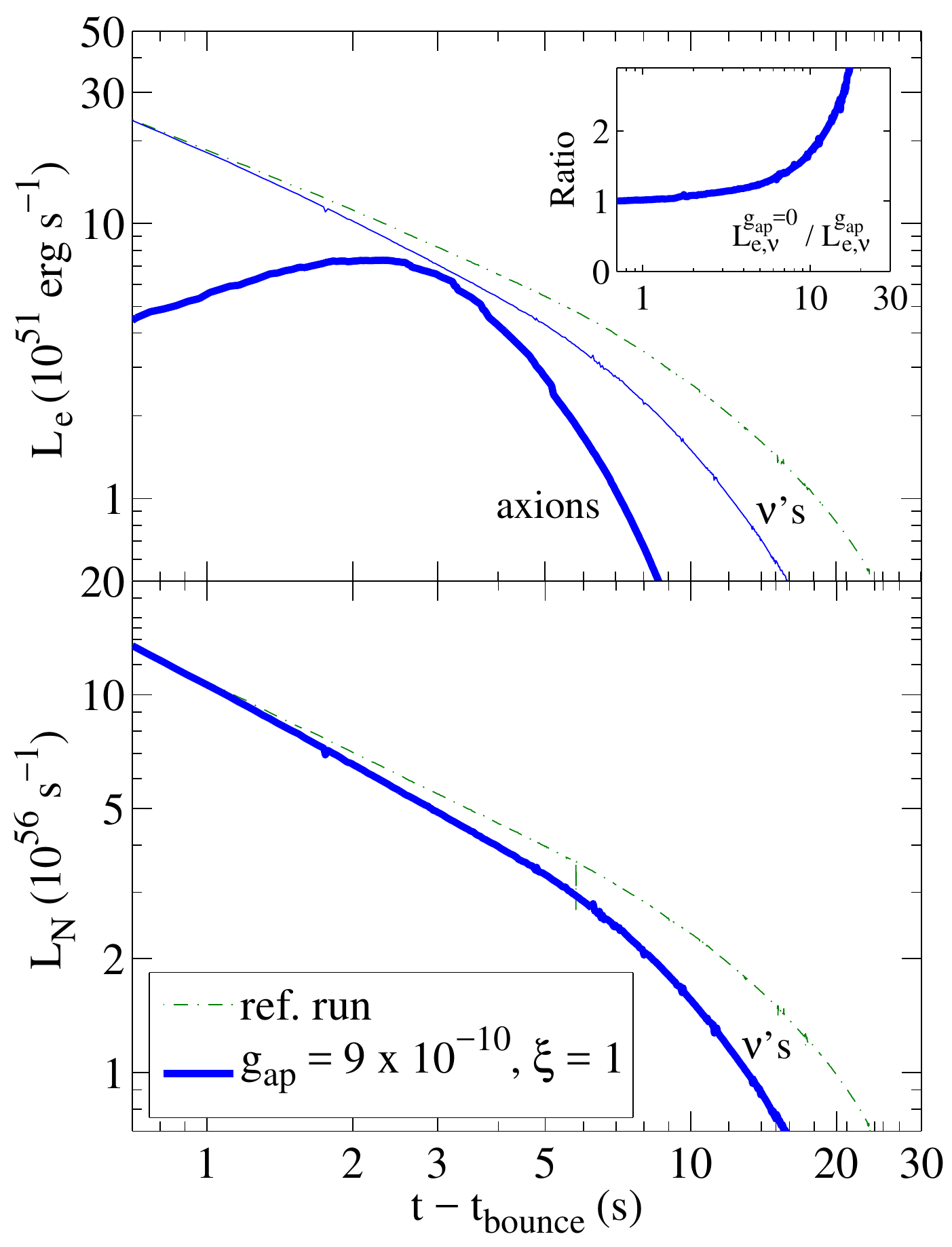}\label{fig:h11c_lumin_gap}}
\\
\subfigure[~Average neutrino energies]{
\includegraphics[width=0.48\textwidth]{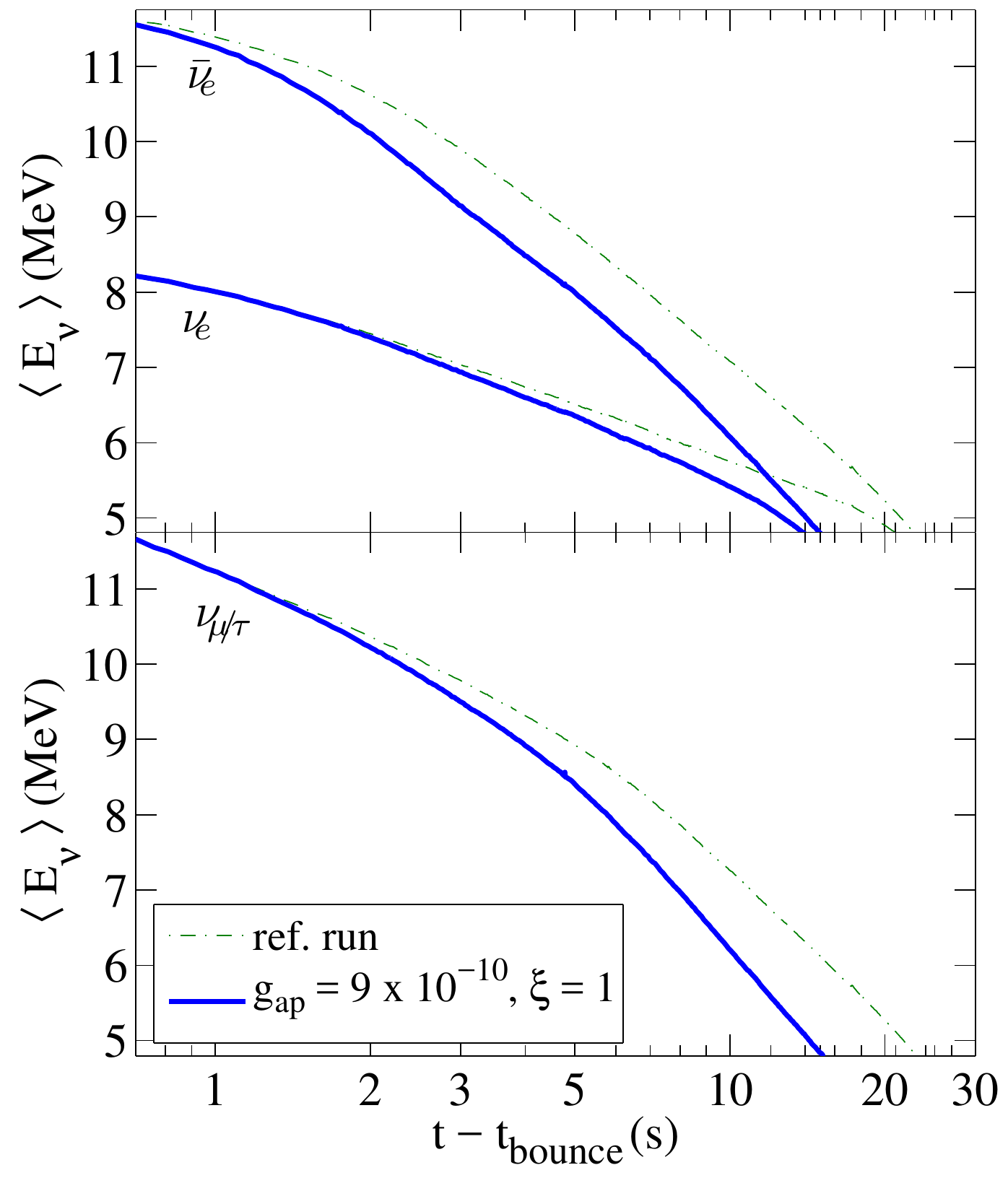}\label{fig:h11c_rms_gap}}
\caption{(color online) Neutrino signal for s11.2, comparing the reference run (green dash-dotted lines) and the simulation with axions with $g_{\rm ap} = 9\times 10^{-10}$ and $\xi=1$ (solid blue lines); same setup as Fig.~\ref{fig:neutrinos_s18}.}
\label{fig:neutrinos_s11.2}
\end{figure}

Moreover, we also explore the impact of variations of the saturation value $\Gamma_\sigma^{\rm max}$ in Fig.~\ref{fig:axion}(c): in addition to our primary choice of 60~MeV we select 30~MeV and 120~MeV as representative lower and upper bounds leading to strong and weak suppression of the axion emission rate towards high density. Correspondingly we find a low (large) axion luminosity for $\Gamma_\sigma^{\rm max}=30 (120)$~MeV in Fig.~\ref{fig:axion}(b).

During the subsequent PNS evolution the axion luminosity rises slowly, as illustrated in the top panel of Fig.~\ref{fig:neutrinos_s18}, according to the PNS contraction accompanied by the rise of the central density and temperature. The axion luminosity reaches a maximum value that corresponds to the decrease of the central temperature, see top panels in Fig.~\ref{fig:hydro}(a), between 1--2~s post bounce. From the comparison with the reference simulation, left panel in Fig.~\ref{fig:hydro}(a), it becomes evident that core cooling starts significantly earlier for the case with additional axion cooling. In particular, the core temperature rise due to the PNS contraction of the reference simulation is never obtained for the simulation with axions. Instead, the gravitational binding energy gain is carried away efficiently by axions instantaneously. The resulting accelerated cooling from axion inclusion leads to core temperatures on the order of $T\simeq 1$~MeV at about 30~s post bounce, while for the reference simulation core temperatures are still in excess of 30~MeV. In Table~\ref{tab:data} we list the values of the maximum luminosities and the corresponding post bounce times, together with the central temperature $T_c$ obtained at $t=10$~s.

The faster cooling and the associated more compact PNS structure shortens the deleptonization timescale, with faster decreasing core neutrino abundances $Y_\nu$ and electron fraction $Y_e$ illustrated in the two bottom panels in Fig.~\ref{fig:hydro}(a). Note that the PNS deleptonization is determined by the decoupling of neutrinos of all flavors at the PNS surface. In particular, high matter temperatures prevent neutrinos from efficient decoupling~\cite{Tubbs:1975jx,Yueh1976a,Yueh:1976b,Yueh:1977b}. Essential therefore is final-state Pauli blocking for electrons (charged-current reaction (1) in Table~\ref{tab:nu-reactions}) and neutrons (neutral-current scattering reaction (4) in Table~\ref{tab:nu-reactions}), for both of which the opacity reduces with increasing temperature. Hence only towards late times, with decreasing temperature, neutrinos can decouple also at high densities. The enhanced cooling for the simulation with axions affects not only the core temperature. The temperature at the PNS surface is also significantly lower than for the reference model. This is a feedback from the faster core contraction. It allows neutrinos to decouple deeper inside the PNS surface at generally higher density. This has important consequences for the neutrino signal, as illustrated in Fig.~\ref{fig:neutrinos_s18}, which results in reduced neutrino energy fluxes ($L_{{\rm e},\nu}$) and number fluxes ($L_{{\rm N},\nu}$) as well as the average energies, in comparison to the reference model; their ratio is shown in the bottom panel of Fig.~\ref{fig:h18a_lumin_gap}. It becomes increasingly important towards late times when PNS structure differences become large, as illustrated in Fig.~\ref{fig:hydro}.

At about 10~s the neutrino luminosity is reduced by a factor of 2 (see Table~\ref{tab:data} and Fig.~\ref{fig:h18a_lumin_gap}) and at 20~s the reduction exceeds one order of magnitude. In Table~\ref{tab:data} we also list the total energy emitted via neutrinos and axions, from which it becomes clear that $E_\nu^{\rm tot}\simeq E_a^{\rm tot}$ for the selected axion emission parameters. We also explored different values of the axion-proton coupling, i.e.\@ $g_{\rm ap}=1-10\times 10^{-10}$. Only for largest values of $g_{\rm ap}$ the energy loss from axion emission is competing with those of neutrinos; for the smallest values $g_{\rm ap}=1-3\times 10^{-10}$ the impact is in fact negligible. In order to illustrate the impact for smaller values of $g_{\rm ap}$ in Fig.~\ref{fig:neutrinos_s18} we also show the neutrino signal for the case ($g_{\rm ap}=6\times 10^{-10},\xi=0$). The reduced axion cooling, in comparison to ($g_{\rm ap}=9\times 10^{-10},\xi=0$), results in somewhat smaller impact on the neutrino signal with slightly less reduced neutrino fluxes and average energies towards late times. The associated losses are summarized in Table~\ref{tab:data} also for this model, with slightly higher core temperature $T_c$ and generally lower axion losses $E_a^{\rm tot}$ compared to the case with $g_{\rm ap}=9\times 10^{-10}$.

Towards later times, the axion emission rate decreases as a consequence of the continuously reducing core temperature due to the strong temperature-dependence of $Q_a$ in Eq.~\eqref{eq:Qa}. Hence the axion luminosity reduces accordingly (see Fig.~\ref{fig:h18a_lumin_gap}) and consequently axions cannot contribute anymore to the cooling. In addition, at about 30~s post bounce we reach temperatures of the order of about $T\sim1$~MeV, where neutrinos decouple basically at all densities. This corresponds to the domain where the transition from neutrino diffusion to freely streaming takes place and where the nuclear medium starts to modify weak processes significantly at densities in excess of nuclear saturation density, e.g., the modified Urca processes start to dominate the further cooling. Since none of them are included into the current simulation setup it is not meaningful to follow the evolution any longer. 

\subsection{Dependence on the stellar model}
In addition to s18 -- with the baryon ($M_{\rm B}^{\rm PNS}$) and gravitational masses ($M_{\rm G}^{\rm PNS}$) of the PNS at the end of our simulations listed in Table~\ref{tab:data}  -- we also consider s11.2 with a significantly lighter PNS (see Table~\ref{tab:data}). The reference simulation of s11.2 without axions has been published in Ref.~\cite{MartinezPinedo:2014}. Comparing these PNS properties with those from the simulations with axions: differences obtained are on the oder of $10^{-4}$ from a slightly different mass ejection associated with the neutrino-driven wind ejected from the PNS surface during the PNS deleptonization.

Simulation results for s11.2 are in qualitative agreement with those of s18, as illustrated in Fig.~\ref{fig:hydro}(b). However, quantitative differences arise in the magnitude of the axion luminosity and associated enhanced PNS cooling, i.e.\@ for the same value of $g_{\rm ap}$ the PNS deleptonization timescale is somewhat less reduced compared to the reference run ($g_{\rm ap}=0$). This is related to the axion emission rate (integrand of Eq.~\eqref{eq:La}) and in particular to the smaller enclosed mass inside the PNS. Moreover, the central densities and the core temperatures are lower compared to the more massive s18, in particular in the region where axions are produced according to Eq.~\eqref{eq:Qa}; see therefore also the peak axion luminosities of s11.2 in Table~\ref{tab:data} as well as the evolution of neutrino and axion luminosities in Fig.~\ref{fig:h11c_lumin_gap} in comparison to s18. This results in a significantly lower total energy loss from axion emission ($E_a^{\rm tot}$) compared to the more massive progenitor model.

The generally less pronounced impact on the PNS evolution for this lighter stellar model results also in a less pronounced impact on the PNS structure as well as on the evolution of neutrino luminosities and average energies. For s11.2 axions carry away less efficiently heat from the their core (see Table~\ref{tab:data}) -- here $E_\nu^{\rm tot}\simeq 3\times E_a^{\rm tot}$. This qualitative feature has already been reported in Ref.~\cite{Keil:1997}. Consequently the impact on the reduced PNS deleptonization timescale is weaker and the reduction of the neutrino fluxes and average energies if less pronounced,  as illustrated in Figs.~\ref{fig:h11c_lumin_gap} and \ref{fig:h11c_rms_gap}.

\begin{figure}[t!]
\includegraphics[width=0.48\textwidth]{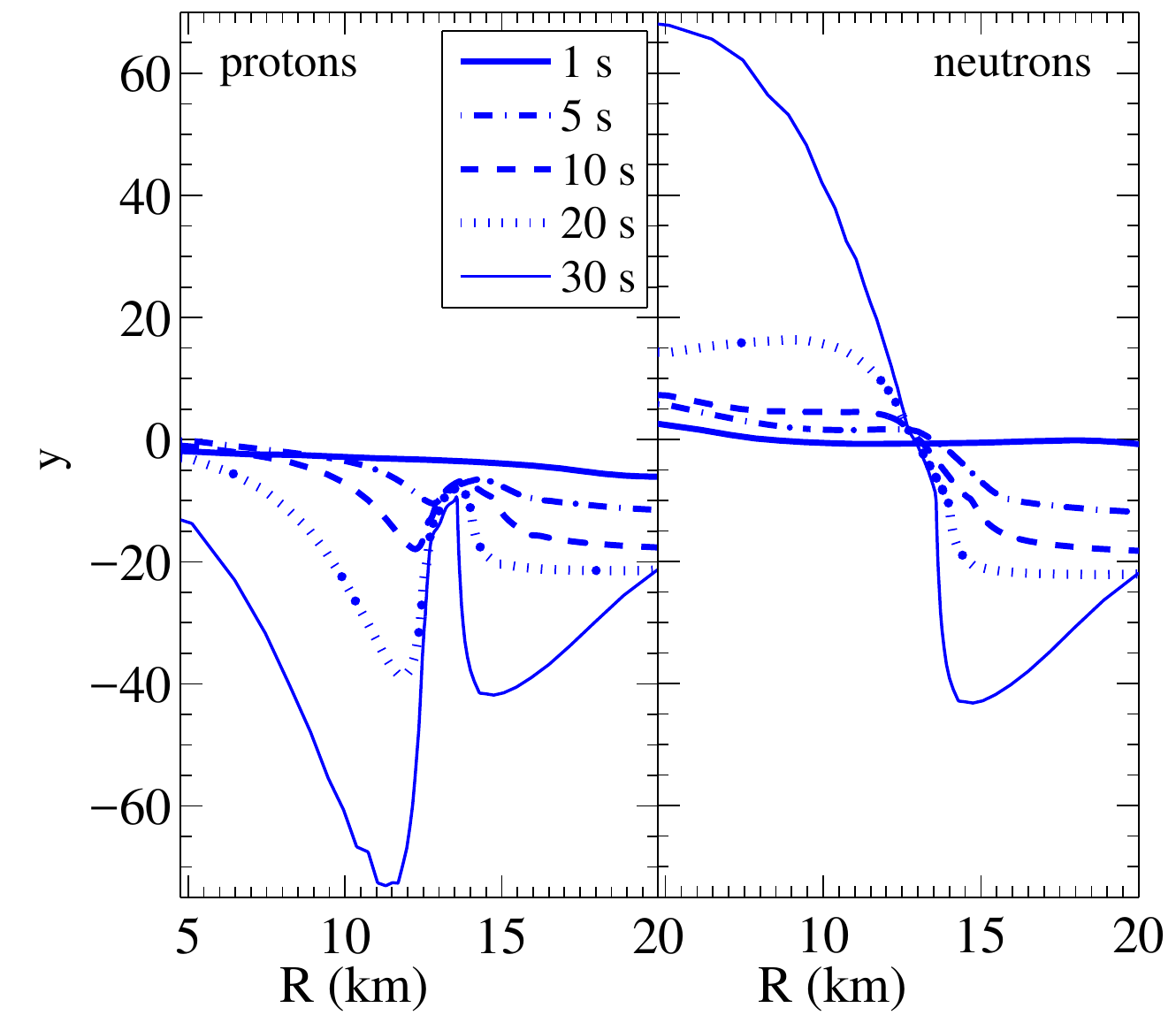}
\caption{(color online) Degeneracy for protons ({\em left panel}) and neutrons ({\em right panel}) for the simulation with axions ($g_{\rm ap}=9\times 10^{-10}$, $\xi=1$) at selected post bounce times corresponding to the PNS profiles illustrated in the right panel of Fig.~\ref{fig:hydro}~(a).}
\label{fig:eta}
\end{figure}

\subsection{Role of degeneracy}
Up to this point we compared and analyzed simulation results with $g_{\rm ap}=9\times 10^{-10}$ and zero proton degeneracy, i.e.\@ $\xi=1$. In order to study the role of degeneracy, we select in addition $\xi=0$ as degenerate limit for the same value of $g_{\rm ap}$ in the simulation of s18 (see Table~\ref{tab:data}). We find, in agreement with the earlier study in Ref.~\cite{Keil:1997}, that for the axion emission degeneracy plays only a marginal role with negligible impact on the overall PNS evolution as well as on the neutrino and axion luminosities (see therefore the red lines in Figs.~\ref{fig:h18a_lumin_gap} and \ref{fig:h18a_rms_gap}). This can be understood since protons, unlike neutrons, are only partly degenerate (if at all, $y=\mu^0/T < 0$). Degenerate and non-degenerate approximations have been compared in Ref.~\cite{Brinkmann:1988vi} (see their Fig.~2) from which it becomes clear that for $y=-1$ both approaches coincide. In support Fig.~\ref{fig:eta} illustrates radial profiles of $y$ at selected post bounce times during the PNS deleptonization corresponding to the conditions shown in Fig.~\ref{fig:hydro}~(a), for protons (left panel) and neutrons (right panel). Note that the nucleon chemical potentials $\mu^0_p$ and $\mu^0_n$ don't include the rest masses.

Towards late times when the temperature decreases neutrons become highly degenerate, however, the proton degeneracy also decreases. Nevertheless late times correspond to conditions when the axion production becomes negligible due to the low temperatures (note again the strong temperature dependence of the axion production rate Eq.~\eqref{eq:Qa}), mildly independent from properties of the contributing protons.  

\section{Impact on the observable neutrino signal}

In this section we will show how the modification of the SN neutrino signal due to the emission of axions would affect the observable neutrino signal in large underground detectors

\subsection{Overview of the calculation}
The  $\nu$ event rate $N_e$ at Earth can be expressed symbolically as follows~\cite{Fogli:2004ff},
\begin{equation}
N_e= F_\nu \otimes \sigma_e \otimes R_e \otimes \varepsilon~,
\end{equation}
where the oscillated $\nu$  flux at Earth is convoluted with the interaction cross section $\sigma_e$ in the detector for the  production of an electron or a positron, as well as the energy-resolution function $R_e$ of the detector and the detection efficiency $\varepsilon$. Since we will always show energy-integrated quantities (e.g.\@ the neutrino light curves) the energy resolution plays no role. Therefore we will neglect its effect. Moreover we assume $\varepsilon=1$  above the threshold. 

\subsubsection{Original neutrino fluxes}
\begin{figure}[t!]
\includegraphics[width=0.48\textwidth]{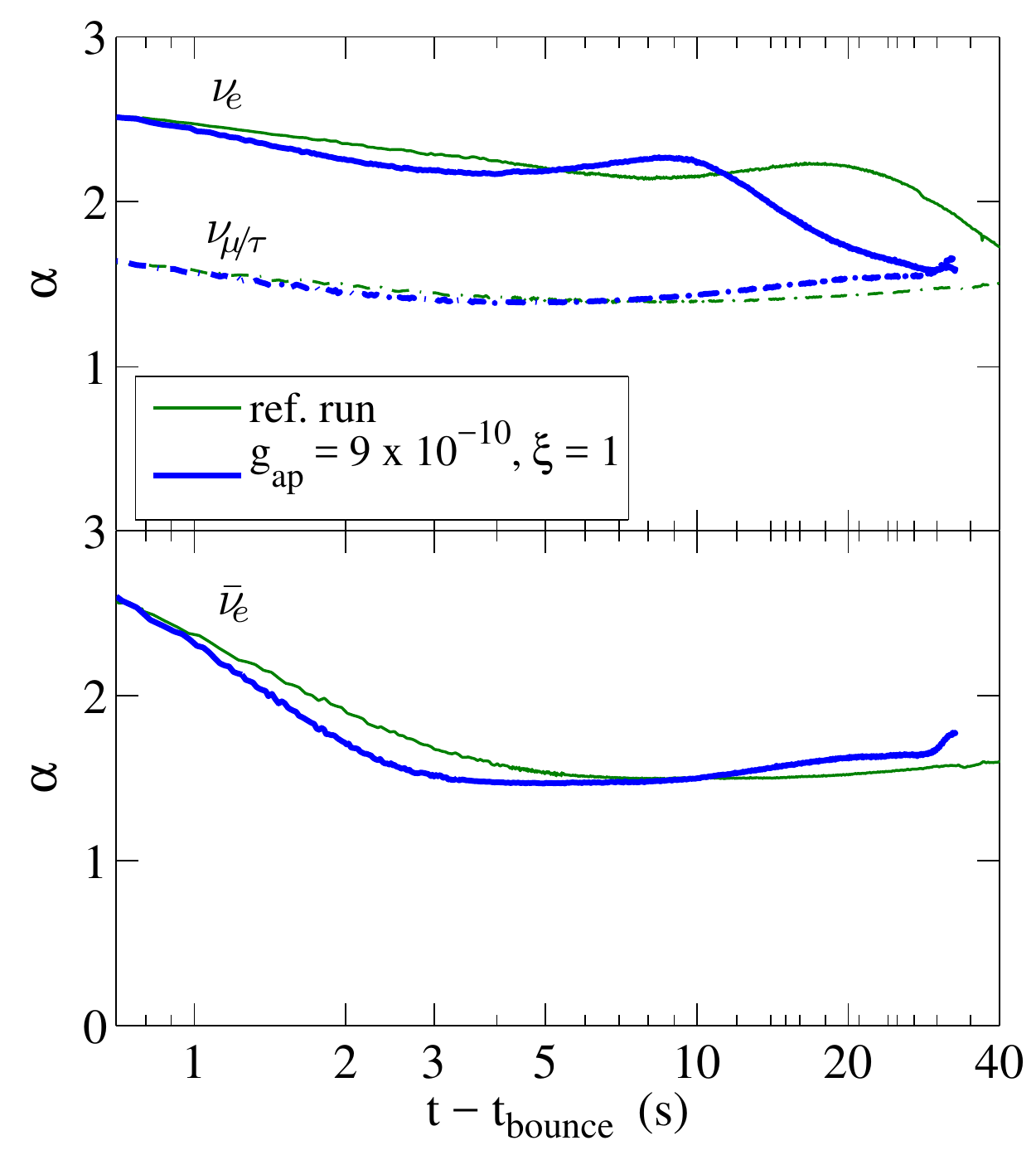}
\caption{(color online) Dimensionless energy-shape parameter Eq.~\eqref{eq:alpha} for s18 comparing the reference case without axions and the simulation with axions ($g_{\rm ap}= 9 \times 10^{-10}, \xi=1$).}
\label{fig:alpha}
\end{figure}

The bare $\nu$ distributions obtained from the supernova simulations, i.e. without neutrino oscillations considered, are parametrized in energy and time as follows,
\begin{equation}
F^0_\nu = \phi_\nu(t) f_\nu(E,t) = \frac{L_{\rm e,\nu}(t)}{\langle E_\nu(t) \rangle} f_\nu(E,t)~,
\end{equation}
with $\nu =\{\nu_e, \bar\nu_e, \nu_x (=\nu_\mu,\nu_\tau)\}$ and where $\phi_\nu(t)$ is the \emph{energy-integrated} neutrino number flux for each post-bounce time $t$ in terms of $L_\nu(t)$ and $\langle E_\nu(t) \rangle$. The function $f_\nu (E,t)$ is the energy spectrum, normalized such that $\int dE f_\nu(E,t) =1$. A useful parameterization of this spectrum is given in terms of a quasi-thermal distribution known as $\alpha$-fit~\cite{Keil:2002in},
\begin{eqnarray}
f_\nu(E,t) &=& \frac{1}{\langle E_\nu(t) \rangle} \frac{(1+\alpha_\nu(t))^{(1+\alpha_\nu(t))}}{\Gamma(1+\alpha_\nu(t))}  \left(\frac{E}{\langle E_\nu(t) \rangle} \right)^{\alpha_\nu(t)}
\nonumber \\
&\times&  \exp\left\{-(1+\alpha_\nu(t))\frac{E}{\langle E_\nu(t) \rangle} \right\}~,
\label{eq:spectrum}
\end{eqnarray}
with the energy-shape parameter $\alpha_\nu(t)$ given as follows,
\begin{equation}
\alpha_\nu(t)= \frac{2 \langle E_\nu(t) \rangle^2 -  \langle E_\nu(t)^2 \rangle}{\langle E_\nu(t)^2 \rangle-\langle E_\nu(t) \rangle^2}~.
\label{eq:alpha}
\end{equation}
It is given in terms of the root-mean square neutrino energies $\langle E^2 \rangle$ and the average neutrino energies $\langle E \rangle$, which in turn are determined via the neutrino transport from the SN simulations (see Fig.~\ref{fig:alpha}).

\begin{figure}[htp!]
\subfigure[~Ref. case ($g_{\rm ap}=0$) -- mass accretion phase]{
\includegraphics[width=0.425\textwidth]{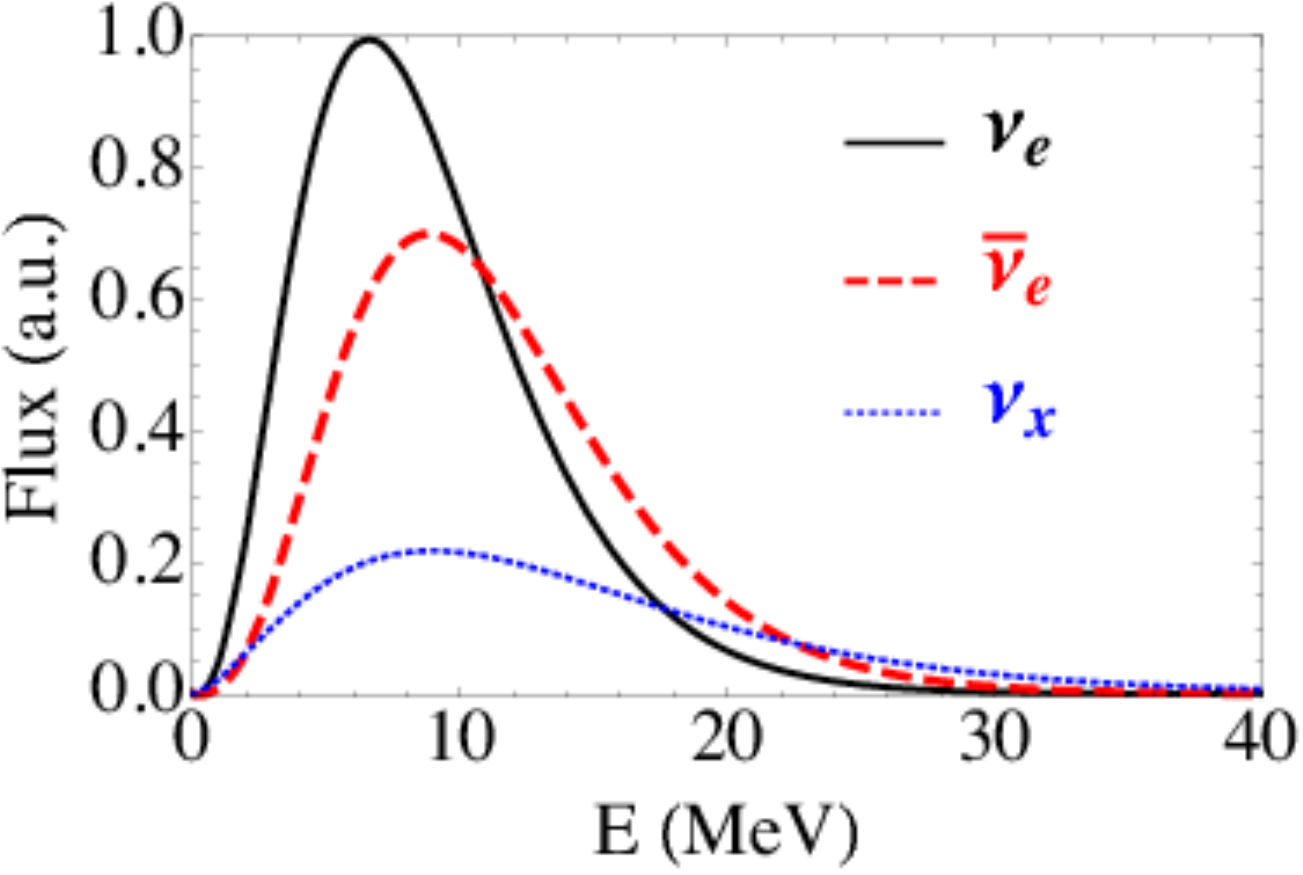}
\label{fig:fluxes_ref_acc}}
\\
\subfigure[~Ref. case ($g_{\rm ap}=0$) -- PNS deleptonization phase]{
\includegraphics[width=0.425\textwidth]{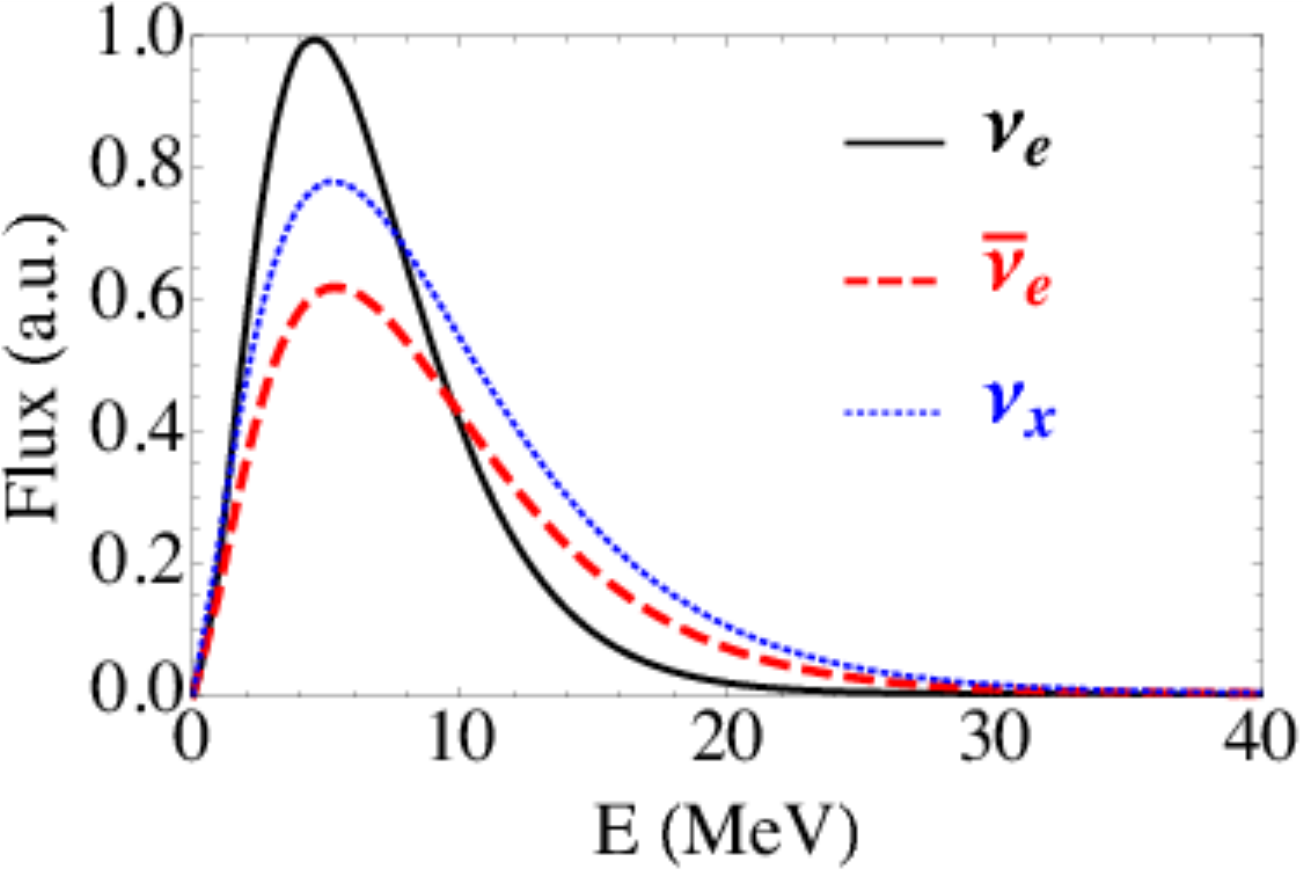}
\label{fig:fluxes_ref_del}}
\\
\subfigure[~($g_{\rm ap}=9\times 10^{-10},\xi=1$) -- PNS deleptonization phase]{
\includegraphics[width=0.425\textwidth]{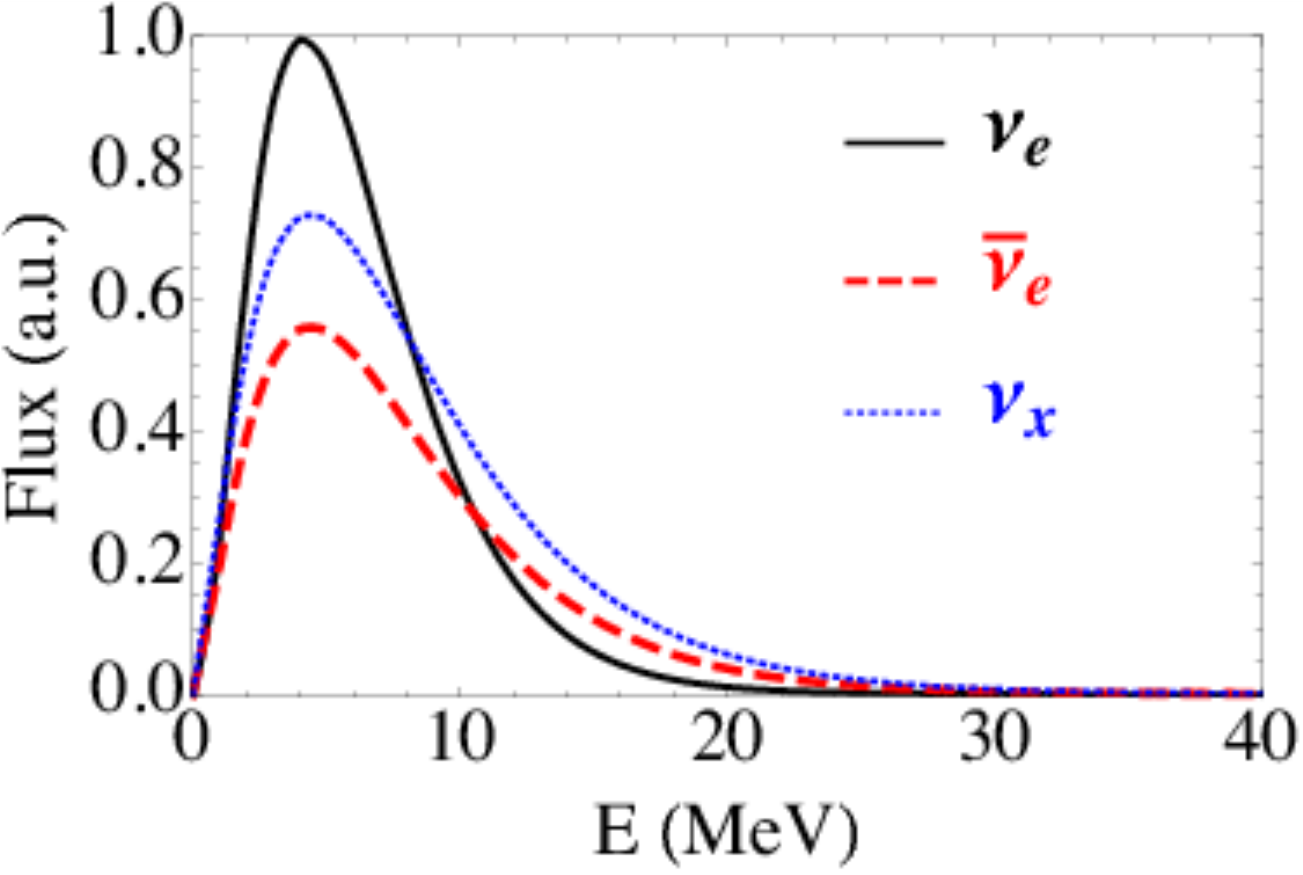}
\label{fig:fluxes_gap9}}
\\
\subfigure[~($g_{\rm ap}=9\times 10^{-10},\xi=1$) -- PNS deleptonization phase]{
\includegraphics[width=0.425\textwidth]{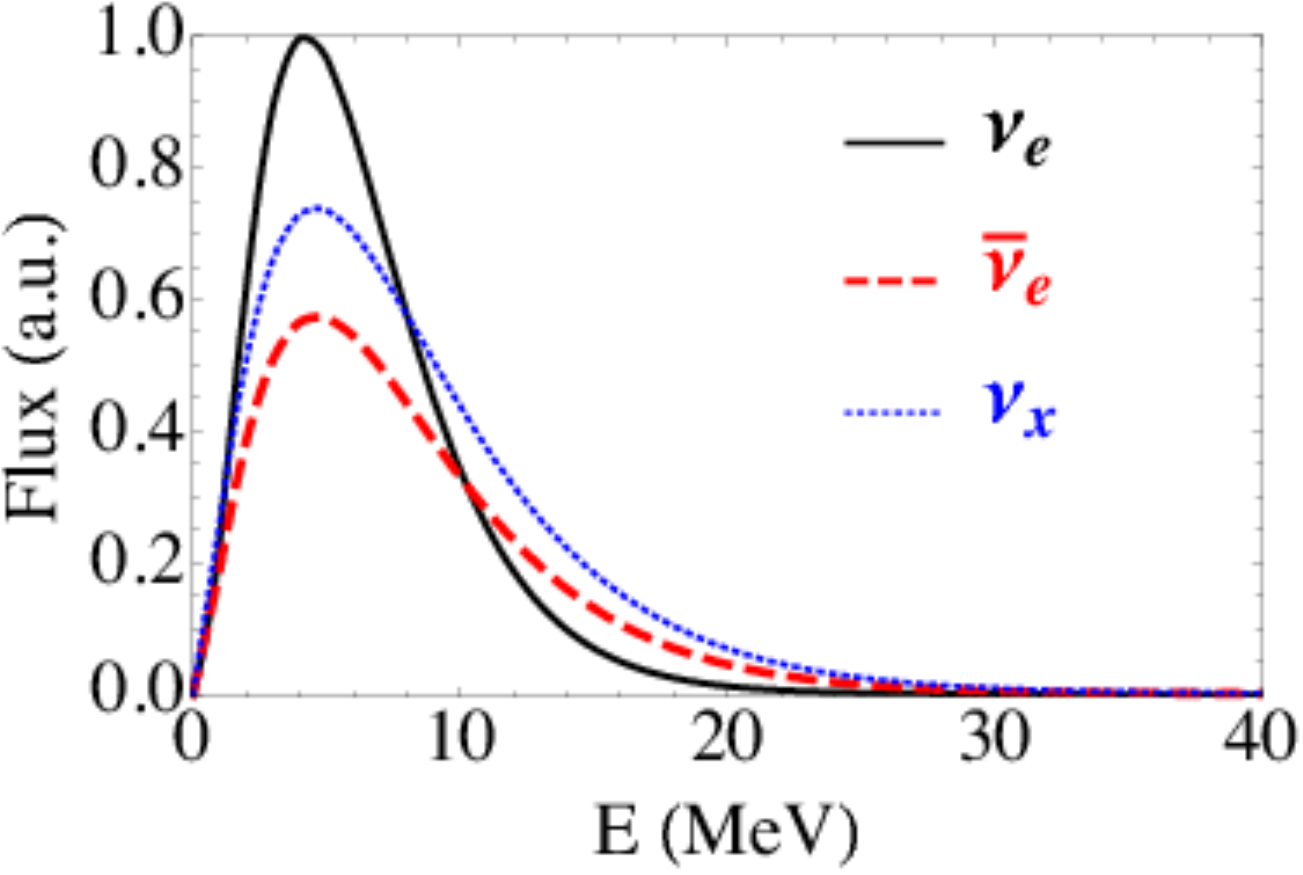}
\label{fig:fluxes_gap6}}
\caption{Time-integrated neutrino  energy spectra for s18.}
\label{fig:fluxes}
\end{figure}

In Fig.~\ref{fig:fluxes} we show the time-integrated $\nu$ energy spectra Eq.~\eqref{eq:spectrum} evaluated at neutrino freeze-out conditions for the different species for s18, separated into accretion phase ($t\in [0,0.3]$~s) in Fig.~\ref{fig:fluxes_ref_acc} and PNS deleptonization phase in Fig.~\ref{fig:fluxes_ref_del}. It is well known that flavor differences among the different neutrino species are large during the accretion phase, while during the deleptonization the neutrino fluxes of different flavors become rather similar (especially in the antineutrino sector) (see sec.~\ref{SNref}). This would diminish the impact of neutrino oscillation effects on the neutrino signal.

\begin{table*}[!t]
\caption{Spectral-fit parameters for the neutrino and antineutrino fluxes for s.18 integrated over  the accretion phase ($t < 0.3$~s) for the reference case ($g_{\rm ap}= 0$) and over the deleptonization phase, comparing the reference case with ($g_{\rm ap}= 6 \times 10^{-10},\xi=0$) and ($g_{\rm ap}= 9 \times 10^{-10},\xi=1$).}
\begin{center}
\begin{tabular}{rlcccccc}
\hline
&  Model/setup &  $\mean{E_{\nu_e}}$ (MeV)  & $\mean{E_{\nu_x}}$ (MeV) & $\Phi_{\nu_e}^0 (10^{56} {\rm s}^{-1})$ & $\Phi_{\nu_x}^0 (10^{56} {\rm s}^{-1})$ & ${\alpha}_{\nu_e}$ &  ${\alpha}_{\nu_x}$ \\
\hline \hline
accretion ($t \le 0.3$~s) & ref. run ($g_{\rm ap}=0$)    &  8.80 & 14.08  & 9.76 & 3.84 & 2.91 & 1.72 \\
deleptonization ($t> 0.3$~s) & ref. run ($g_{\rm ap}=0$)     &  6.65 & 9.05 & 8.73 & 10.06 & 2.15 & 1.38 \\
deleptonization ($t> 0.3$~s) & ($g_{\rm ap}=6 \times 10^{-10},\xi=0$)     &  6.22 & 8.20 &7.58 & 8.00 & 2.12 & 1.31 \\
deleptonization ($t> 0.3$~s) & ($g_{\rm ap}=9 \times 10^{-10},\xi=1$)     &  6.11 & 7.91 &7.18 & 7.39 & 2.12 & 1.29 \\
\hline
\\
\hline
&  Model/setup &  $\mean{E_{\bar\nu_e}}$ (MeV)  & $\mean{E_{\bar\nu_x}}$ (MeV)  & $\Phi_{\bar\nu_e}^0$ $(10^{56} {\rm s}^{-1})$ & $\Phi_{\bar\nu_x}^0 $ $(10^{56} {\rm s}^{-1}$) & ${\alpha}_{\bar\nu_e}$ & ${\alpha}_{\bar\nu_x}$ \\
\hline
\hline
accretion ($t \le 0.3$~s) & ref. run ($g_{\rm ap}=0$)    &  11.27 & 14.08  & 8.32 & 3.84 & 3.51 & 1.72 \\
deleptonization ($t> 0.3$~s) & ref. run ($g_{\rm ap}=0$)     &  8.82 & 9.05 & 7.66 & 10.06 & 1.54 & 1.38 \\
deleptonization ($t> 0.3$~s) & ($g_{\rm ap}=6 \times 10^{-10},\xi=0$)     &  7.96 & 8.20 & 5.99 & 8.00 & 1.41 & 1.31 \\
deleptonization ($t> 0.3$~s) & ($g_{\rm ap}=9 \times 10^{-10},\xi=1$)     &  7.69 & 7.91 &5.45 & 7.39 & 1.38 & 1.29 \\
\hline
\end{tabular}
\end{center}
\label{tab:fluxes}
\end{table*}

In Figs.~\ref{fig:fluxes_gap9} and \ref{fig:fluxes_gap6} we consider the corresponding integrated spectra during the deleptonization phase in the presence of axion emission for $g_{\rm ap} = 9 \times 10^{-10}$ ($m_a \simeq 3 \times 10^{-2}$~eV, $f_a \simeq 4.8 \times 10^{8}$~GeV) and $g_{\rm ap} = 6 \times 10^{-10}$ ($m_a \simeq 8 \times 10^{-3}$~eV, $f_a \simeq 7.3 \times 10^{8}$~GeV) respectively. It becomes evident that for the models with axion emission the spectra are shifted towards lower energies with respect to the reference case. Based on the present description of axion emission, their spectra remain unknown. Nevertheless, they could be extracted directly from the SN simulation following, e.g., Eq.~(8) of Ref.~\cite{Raffelt:2011ft}. This would result in average axion energies far in excess of the average neutrino energies, since axions are produced in hotter and deeper SN regions. In Table~\ref{tab:fluxes} we report the parameters of the time-integrated SN neutrino spectra for the simulations distinguishing between the accretion ($t\leq 0.3$~s) and the deleptonization phase ($t>0.3$~s).

\subsubsection{Flavor conversions}

The initial neutrino distributions are in general modified by flavor conversions $F^0_\nu \to F_\nu$. We assume a standard $3\nu$ framework where the mass spectrum of neutrinos is parameterized in terms of two mass-squared differences, whose
values are obtained from a $3\nu$ global analysis of the neutrino data~\cite{Capozzi:2016rtj},
\begin{eqnarray}
\Delta m^2_{\rm atm} &=& m_3^2 -m_{1,2}^2 = 2.50 \times 10^{-3} \textrm{eV}^2~,\\
\Delta m^2_{\rm \odot} &=&  m_2^2 -m_1^2 = 7.37 \times 10^{-5} \textrm{eV}^2~,
\end{eqnarray}
where according to the sign of $\Delta m^2_{\rm atm}$ one distinguishes a normal (NH,  $\Delta m^2_{\rm atm}>0$) or an inverted (IH,  $\Delta m^2_{\rm atm}<0$) mass ordering. The flavor eigenstates $\nu_{e,\mu,\tau}$ are a linear combination of the mass eigenstates $\nu_{1,2,3}$ by means of three mixing angles. Their best-fit values according to the global analysis are (for the NH case),
\begin{eqnarray}
\sin^2 \theta_{12} = 0.297~, \;\;\; \sin^2 \theta_{13} = 0.0214~.
\end{eqnarray}
The value of the mixing angle $\theta_{23}$ is not relevant in our context since we are assuming equal $\nu_\mu$ and $\nu_\tau$ fluxes. For IH case the best-fit values are similar to the ones quoted before.

Neutrino flavor conversions in SNe are a fascinating and complex phenomenon where different effects would contribute to profoundly modify the original neutrino fluxes (see Ref.~\cite{Mirizzi:2015eza} for a recent review). Indeed, in the deepest SN regions ($r \lesssim 10^3$~km) the neutrino density is sufficiently high to produce a self-induced refractive term for the neutrino propagation, associated with $\nu$--$\nu$ interactions. These would produce surprising collective effects in the flavor dynamics that are currently under investigation~\cite{Chakraborty:2016yeg}. At larger radii ($r \sim 10^4$--$10^5$~km) neutrino fluxes would be further processed by the ordinary Mikheeyev-Smirnov-Wolfenstein (MSW)  matter effects~\cite{Wolfenstein:1977ue,Mikheev:1986gs}. The sensitivity of the matter effects to the SN dynamics has been discussed in the literature; notably  concerning the shock-wave propagation and the matter density fluctuations. Furthermore, if neutrinos cross the Earth before their detection, this could induce additional oscillation effects. In the following, we neglect all these complications, since the main signature of the axion emission would be the overall reduction of the cooling time on the $\nu$ light curve. Oscillation effects would be sub-leading (cf. Ref.~\cite{Wu:2015}). Moreover, from Figs.~\ref{fig:neutrinos_s18} and \ref{fig:neutrinos_s11.2} one realizes the spectral differences among the different $\nu$ species are reduced at $t\gtrsim 5$~s when axion emission plays a major role. Therefore we simply assume that neutrino fluxes can only undergo the traditional MSW flavor conversions along a static density profile. In this case the dependence on $\theta_{13}$ of the flavor conversions disappears. The oscillated $\bar\nu_e$ flux that we consider for the detection is decomposed as follows~\cite{Borriello:2012zc},
\begin{eqnarray}
&{\rm NH}:& F_{\bar\nu_e} = \cos^2 \theta_{12} F^0_{\bar\nu_e} +  \sin^2 \theta_{12} F^0_{\bar\nu_x}~, \\
&{\rm IH}:& F_{\bar\nu_e} =  F^0_{\bar\nu_x}~.
\end{eqnarray}
In the following, for definitiveness we will show our results only in the NH case.

\subsubsection{Neutrino detection}
There are several experiments which aim at detecting SN neutrinos with a high statistics (see Ref.~\cite{Mirizzi:2015eza} for a list of current and future experiments). The presently largest underground detectors with the necessary sensitive to observe SN neutrinos are the water-Cherenkov Super-Kamiokande and the Cherenkov experiment in antartic ice IceCube. These are  mostly sensitive to electron antineutrinos through the inverse beta decay process, ${\bar\nu_e}+p \to n+ e^+$. Moreover, a megatonne water-Cherenkov detector is a realistic future possibility in view of current efforts towards precision long-baseline oscillation experiments. We consider these three detectors as references in our study. For the inverse-beta-decay process, we take the differential cross section from Ref.~\cite{Strumia:2003zx}. The total cross section grows approximatively as $E^2$. For Super-Kamiokande we take a 22.5-kton fiducial mass, while for a future megatonne Cherenkov detector, we assume 400~kton.

\begin{figure*}[t!]
\includegraphics[width=0.9\textwidth]{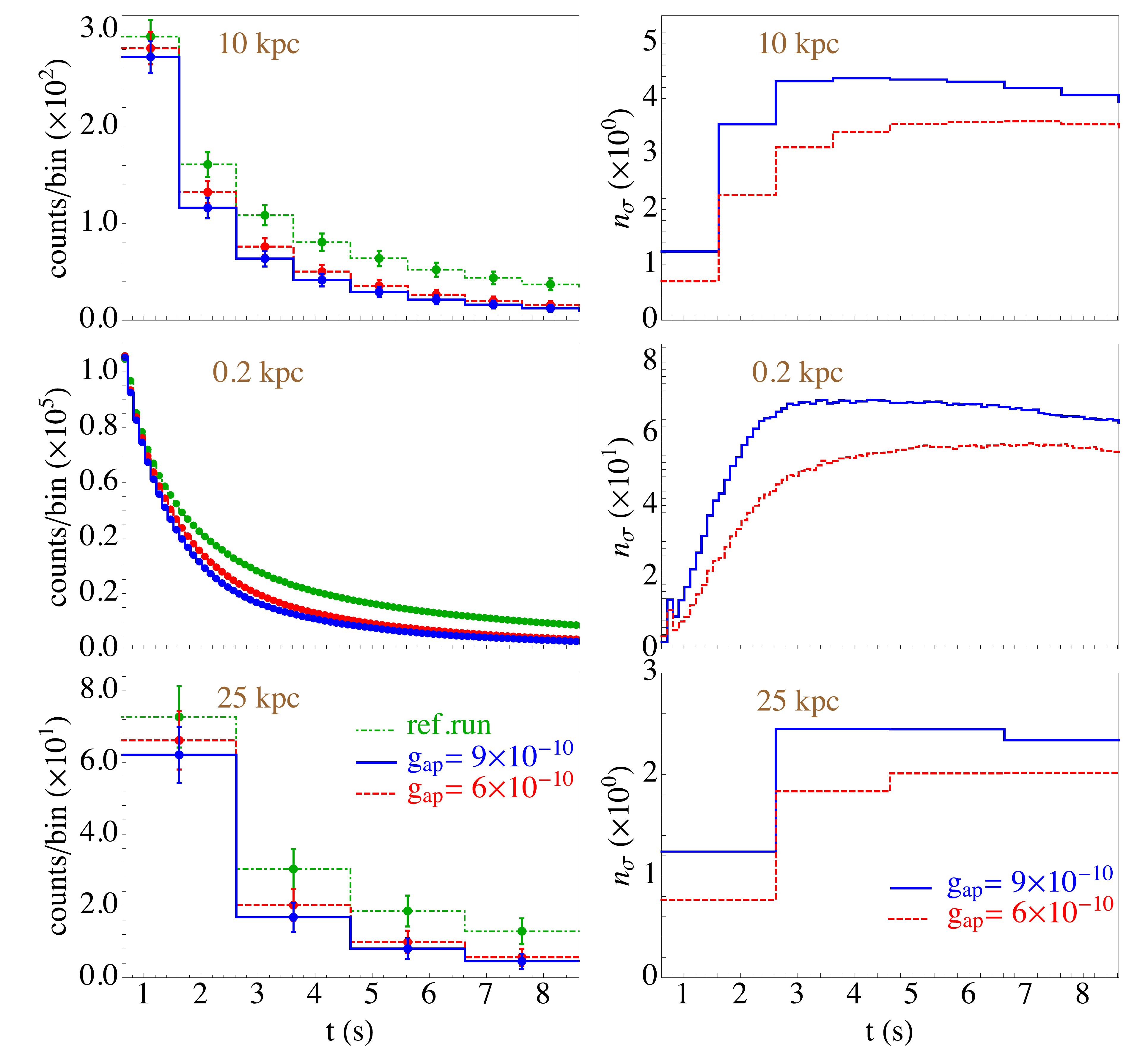}
\caption{(color online) {\em Left panel:} Super-Kamiokande neutrino event rate based on s18 at selected SN distance -- 10~kpc (top panel), 0.2~kpc (middle panel) and 25~kpc (bottom panel) -- comparing the reference model and simulations with axions included ($g_{\rm ap}= 9 \times 10^{-10}, \xi=1$) and ($g_{\rm ap}= 6 \times 10^{-10},\xi=0$). {\em Right panel:} Corresponding ratios relative to the reference case.}
\label{fig:SuperK}
\label{fig:SK}
\end{figure*}

\begin{figure*}[t!]
\includegraphics[width=0.9\textwidth]{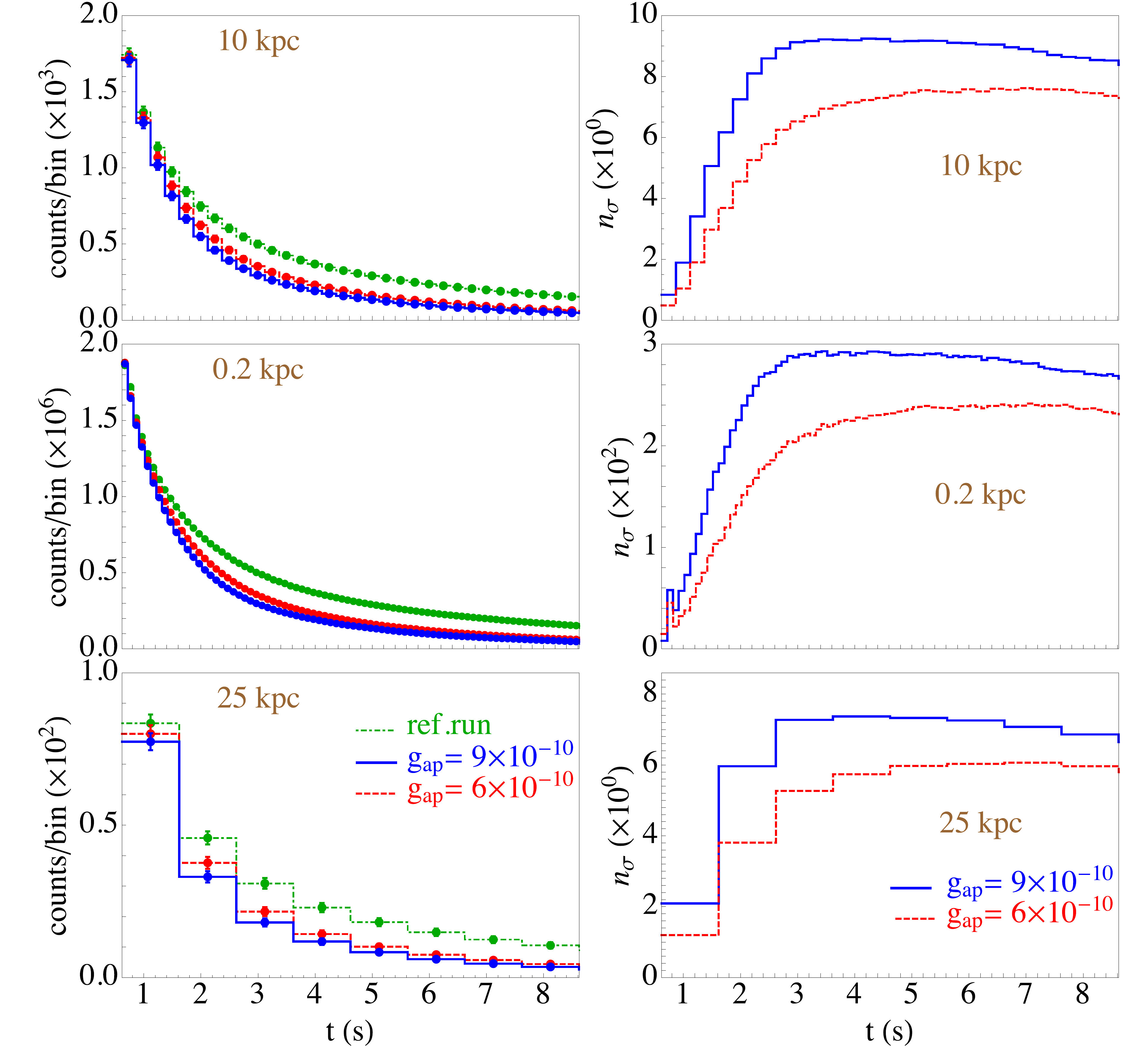}
\caption{(color online) Same as Fig.~\ref{fig:SK} but for the 400  kton Water Chereknov detector.}
\label{fig:Mton}
\label{fig:MT}
\end{figure*}

\begin{figure*}[t!]
\includegraphics[width=0.9\textwidth]{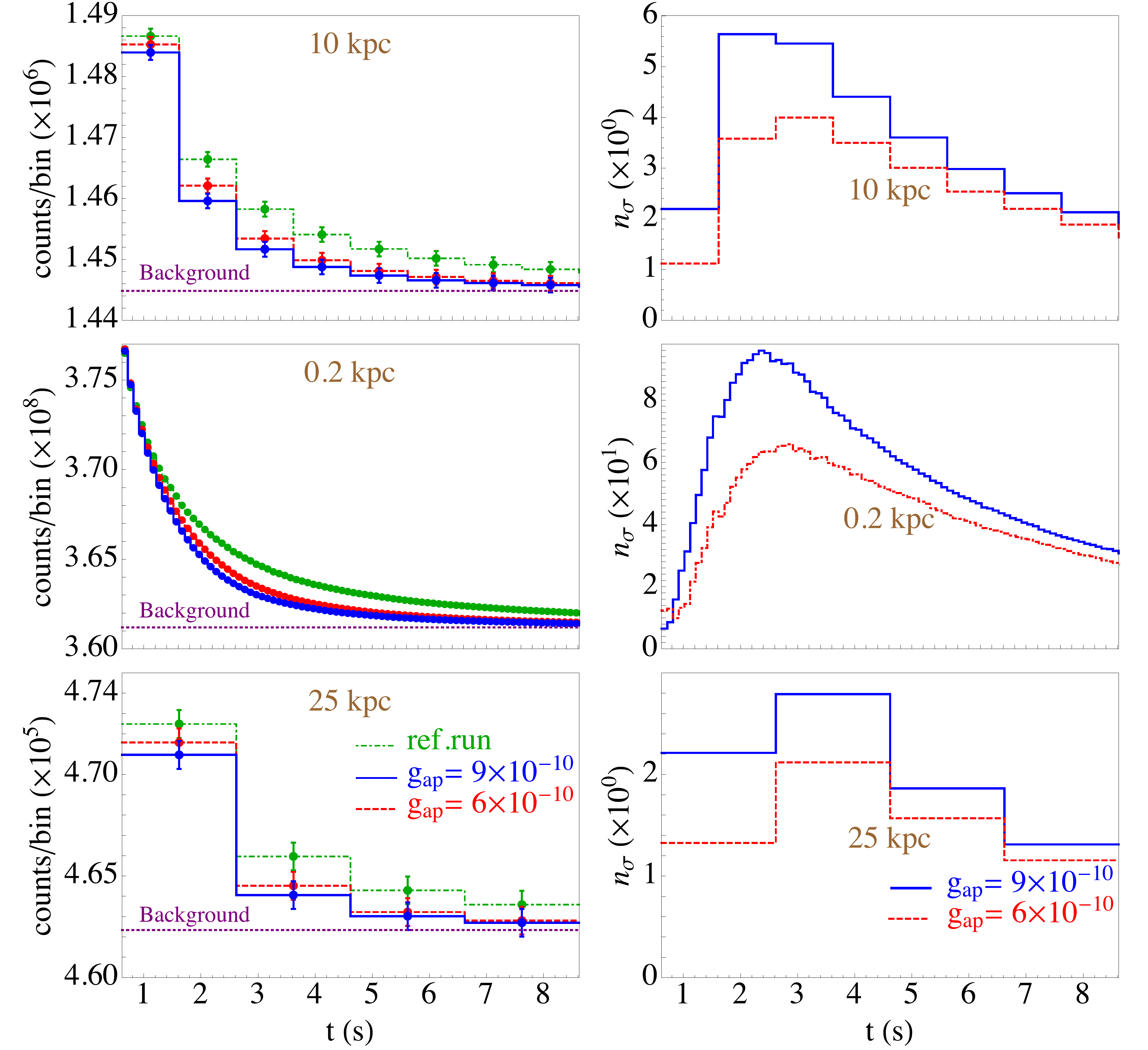}
\caption{(color online) Same as Fig.~\ref{fig:SK} but for the IceCube detector.}
\label{fig:ice}
\label{fig:IC}
\end{figure*}

A galactic-SN $\nu$ burst would be detectable in Icecube by a sudden, correlated increase in the photomultiplier count rate on a timescale of the order of 10~s (see Ref.~\cite{Abbasi:2011ss} for a recent description). In its complete configuration and with its data acquisition system, IceCube has 5160 optical modules~\cite{Abbasi:2011ss} and an effective detection volume of about 3 Mton.  For this reason it represents the largest running detector for SN neutrinos.  The reaction process in the antartic ice would be the inverse beta decay.  However, Icecube being a coarse-grained detector would only pick up the average Cherenkov glow of the ice, being unable of reconstructing the signal on an event-by-event  basis like a water-Cherenkov detector. The detection rate is given as follows~\cite{Dighe:2003be,Serpico:2011ir},
\begin{equation}
{\mathcal R}_{\bar\nu_e} = \int_{0}^{\infty} dE F_{\bar\nu_e} E_{\rm rel} (E) \sigma_e(E)~,
\end{equation}
where $E_{\rm rel} (E)$ is the energy released by a neutrino of energy $E$ and $\sigma_e(E)$ is the cross section for the inverse beta decay process. All other detector parameters (angular acceptance range, average quantum efficiency, number of useful Cherenkov photons per deposited neutrino energy unit, average lifetime of Cherenkov photons, effective photo-cathode detection area) have been fixed to the fiducial values adopted in Ref.~\cite{Dighe:2003be,Serpico:2011ir}, to which we refer to for further details.

We also remind the reader that if an efficient $\nu_e$ detector such as a large liquid Argon Time Projection Chamber becomes available, it would have unique capabilities for reconstructing the $\nu_e$ light curve~\cite{Mirizzi:2015eza}. The axion effect which we will now show on the $\bar\nu_e$ signal would be similar for the $\nu_e$ burst. 

\subsection{Axion impact on the ${\bar\nu_e}$ light curve}
In Fig.~\ref{fig:SK} we show the $\bar\nu_e$ light curve during the deleptonization phase simulated for Super-Kamiokande  for the fiducial case of s18 assuming different distances for the SN to occur: in the Galactic Center at $d= 10$~kpc (upper panel), for the lucky case of a close-by SN at 0.2~kpc (central panel), and the pessimistic case of $d=25$~kpc (lower panel). For the case with  $d= 10$~kpc we use 1~s time bins, for $d= 0.2$~kpc we choose 100~ms time bins, and for $d= 25$~kpc the time binning is 2~s. We compare the reference run (continuous curve) with the case of axion emission with ($g_{\rm ap}= 9 \times 10^{-10},\xi=1$) and ($g_{\rm ap}= 6 \times 10^{-10},\xi=0)$. The counts in each bin follow a Gaussian distribution, with a mean given by the observed number of events and a standard deviation (indicating the 68\% confidence level) given by $\sigma=\sqrt{N}$. This latter is also plotted as vertical bars in each time bin. We see that the two cases with different $g_{\rm ap}$ produce rather close light curves, while the deviation with respect to the standard expectation becomes pronounced at $t \gtrsim 2$~s.

In order to compare the standard case and the case in the presence of axion one has to perform a statistical test under the null hypothesis that there is no relation between the  two distributions. At this regard one has to calculate the $p$-value, which is the probability of observing an effect given that the null hypothesis is true. Statistical  significance  is attained when a $p$-value is less than a given significance level. Assuming that the distributions in the two cases follow Gaussian statistics, the significance level can be expressed in terms of number of standard deviations $n_\sigma$. Discrepancy in the distributions at the level of $2-3\sigma$ indicates a possible hint of axions. At this regard, in order to quantify the difference between the standard case and the presence of axion emission in Fig.~\ref{fig:SuperK} we plot  it in terms of the number of standard deviations $n_\sigma$. We realize that in the case of a SN at  $d= 10$~kpc  the difference can be as large as   $\sim 3 \sigma$ for the case with ($g_{\rm ap}= 6 \times 10^{-10},\xi=0$) (dot-dashed curve) and reach  $\sim 4 \sigma$ for ($g_{\rm ap}= 9 \times 10^{-10},\xi=1$) (dashed curve). Such notable effect cannot be mimicked by other known effects. Notice, however, that in order to claim a possible hint of axion, other effects need to be taken into account. In particular, the impact of the neutron star mass and of the nuclear EoS on the neutrino cooling time need to be investigated. In the case of the explosion of a close-by SN, like  Betelgeuse and Antares (at $d \lesssim 0.2$~kpc) we see that the statistical significance in the case of axion emission would be spectacular. Conversely, for a distant SN  $d=25$~kpc the difference would be at most $\sim 2 \sigma$, preventing us from having a robust hint of axion emission.

In order to show the physics potential for axion emission of a future Mton class water Cherenkov detector, in Fig.~\ref{fig:MT} we show the $\bar\nu_e$ light curve during for a 400~kton water Cherenkov detector in the same format as in Fig.~\ref{fig:SK}. Due to the remarkable improvement in the statistics we used a narrow time binning. In particular, for the case with $d= 10$~kpc we use a 0.25~s time bins, for $d= 0.2$~kpc we select 0.1~ms time bins, and for $d= 25$~kpc the time binning is 1~s. We realize that the improvement with respect to Super-Kamiokande is impressive. In particular, the effect of axion emission would be always distinguishable from the standard expectations also for a distant galactic SN at more than $5 \sigma$
as shown in the right panel of Fig.~\ref{fig:Mton}. 

We comment here that in the case of an extragalactic SN explosion within 1~Mpc, a Mton class detector would collect $\mathcal{O}(10)$ events~\cite{Kistler:2011}. It would be comparable to the neutrino signal detected from SN1987A. In this case one could not perform a detailed study of the $\nu$ light-curve, like the one presented above. However, from the comparison of the total number of events and from the duration of the burst with the expectation from different SN models one would potentially confirm the SN1987A results.

Finally in Fig.~\ref{fig:IC}, we present the event rate of Icecube. The average value of the photomultiplier background noise is represented as horizontal short-dotted curve, with typical error estimates of 280~s$^{-1}$  in each optical module~\cite{Serpico:2011ir}. We realize that also this detector has good capabilities to detect the effect of an axion extra-cooling on the ${\bar\nu}_e$ light curve. For the case with $d= 10$~kpc we use a 1~s time bin, for $d= 0.2$~kpc we apply 100~ms time bins, and for
$d= 25$~kpc the time binning is 2~s. Remarkably, from the right panel of Fig.~\ref{fig:IC} one sees that for a SN at $d= 10$~kpc in case of ($g_{\rm ap}= 9 \times 10^{-10},\xi=1)$ the difference with respect to the standard case is at more than $5 \sigma$ and for ($g_{\rm ap}= 6 \times 10^{-10},\xi=0)$ it is at $4 \sigma$ level. Only in the case of a faraway SN at  $d= 25$~kpc the axion effect is below $3 \sigma$. In this sense Icecube and Super-Kamiokande have similar capabilities.

\section{Summary and conclusions} 

In this survey we review the impact of axion emission in core-collapse SNe from $N$--$N$ bremsstrahlung. The process acts as additional sink and contributes to the cooling of the nascent PNS, which is the central objects in core collapse SNe. PNSs are initially hot and lepton rich, they deleptonize via the emission of neutrinos of all flavors once the supernova explosion has been launched. Unlike neutrinos, which decouple mainly at the PNS surface, axions originate from the PNS interior. This is due to the strong temperature dependence of the axion production rate. Moreover, the axion emission rate is proportional to the number density of nucleons. Hence the local production rate drops to zero towards the PNS surface primarily with decreasing temperature and also with decreasing proton abundance. Note that throughout this study we only consider axion-proton coupling (with finite $g_{\rm ap}$) neglecting axion-neutron coupling.

We implement the associated cooling process in simulations of the PNS deleptonization. We confirm a correlation between $g_{\rm ap}$ and deleptonization timescale, i.e. large(small) values of $g_{\rm ap}$ result in fast(slow) deleptonization of the PNS. The magnitude of the shortened deleptonization depends on the value of $g_{\rm ap}$. Axion emission carries away heat efficiently from the PNS interior which results in a generally more compact structure and lower temperatures in comparison to the reference case, so that neutrinos decouple deeper inside of the PNS surface layer at higher density with lower fluxes and smaller average energies. Our findings are in qualitative agreement with previous studies~\cite{Keil:1997}. For the smallest values of $g_{\rm ap}$ explored here ($g_{\rm ap}=1-1.5\times 10^{-10}$) the impact on the neutrino fluxes and average neutrino energies as well as their evolution is negligible; for largest values ($g_{\rm ap}=6-10\times 10^{-10}$) we find a significant shortening of the PNS deleptonization and a reduction of the associated timescale of neutrino emission, however, still in agreement with SN1987A. From our sensitivity study we find  that values of the axion mass  $m_a \gtrsim 8 \times 10^{-3}$~eV can be probed from a future SN explosion. We stress that this value has to be taken as indicative. Indeed, in order to obtain a sharp bound one has to perform  an appropriate statistical test comparing different SN models  and account for possible  effects that can impact the neutrino cooling time.

The suppression of axion emission due to many-body effects towards increasing density may be important. We treat this via the saturation of the lowest-order effective spin fluctuation rate. Comparing our results with those of the parametric study of axion emission of Ref.~\cite{Keil:1997} (see their Fig.~6) -- the authors focused mainly on simulation results obtained without saturation -- we find that for the same value of the axion-proton coupling the reduction of total neutrino energy loss is significantly smaller, up to a factor 2 when saturation effects are included. Currently large uncertainties regarding the nuclear medium at supersaturation density and at high temperatures prevents us to predict quantitatively the suppression of axion emission due to many-body effects. Even chiral-effective field theory as ab-initio approach to describe dilute neutron matter is applicable only up to normal nuclear matter density, and hence cannot provide further constraints~\cite{Hebeler:2010a,Steiner:2012,Tews:2013}. Such state of matter may be accessible in future heavy-ion collider facilities, e.g., FAIR at the GSI in Darmstadt (Germany) and NICA in Dubna (Russia). However, with a better understanding of the axion-nucleon coupling, it may be possible to determine the magnitude of many-body effect from the detection of the neutrino signal of the next Galactic SN explosion. Note that the generally weak axion losses from low-mass PNSs make it only possible to deduce such analysis for SN explosions of massive or at least intermediate-mass progenitor stars, with typical neutrino losses on the order of $2.5-3.0\times 10^{53}$~erg confirmed by the neutrino detection of SN1987A. Then, neutrino losses significantly below this range would point to additional losses, e.g., axions with large matter coupling and/or small suppression due to weak many-body effects. 

Moreover, we explored the neutrino signal in currently operating and future planned underground neutrino detectors for galactic events, with significant reduction of the event rate due to the emission of axions observable. From the magnitude of suppression it is in principle possible to deduce axion parameters, e.g., mass and couplings. This requires ''good'' supernova models with reliable predictions for the neutrino fluxes and spectra as well as their evolution, in particular for the PNS deleptonization phase from which most neutrinos will be detected. Therefore, the accurate treatment of neutrino transport, e.g., based on Boltzmann transport or in the diffusion limit is essential. At this regards we have shown that the statistics will not be a limiting issue for a typical SN. Currently large uncertainties originate from the unknown super-saturation density EoS, which affects not only the PNS evolution with fast(slow) contraction for soft(stiff) EoS~\cite{Fischer:2016a} but also medium modifications, e.g., correlations which modify weak interactions~\cite{Yakovlev:2001,Blaschke:2004}. All these aspects go beyond our present study and require further investigations.  

We conclude mentioning that the physics potential of a SN neutrino observation is complementary to the reach of planned ALP searches, particularly the International Axion Observatory (IAXO) searching for conversions in photons of axions coming from the Sun~\cite{helioscope:2011}. We remind the reader that IAXO is sensitive to generic axion-like particles coupled to photons and has the potential to probe the QCD axion region up to masses $m_a \gtrsim 10^{-2}$~eV~\cite{2013arXiv1311.0029E}. We have seen that in principle with a galactic SN one could probe also smaller values of the mass. Conversely, if an axion signal were to be found by IAXO, this would change the current SN picture. An axion emission would strongly modify the emitted neutrino fluxes and have impact on the diffuse neutrino background and on the stellar nucleosynthesis.

\section*{Acknowledgements}
The authors thank Irene Tamborra for useful discussions. The work of A.M. is supported by the Italian Ministero dell'Istruzione, Universit\`a e Ricerca (MIUR) and Istituto Nazionale di Fisica Nucleare (INFN) through the ``Theoretical Astroparticle Physics'' projects. The work of A.P. was supported by the German Science Foundation (DFG) within the Collaborative Research Center SFB 676 ``Particles, Strings and the Early Universe.'' T.F. acknowledges support from the Polish National Science Center (NCN) under grant number DEC-2011/02/A/ST2/00306. S.C. acknowledges partial support by the Deutsche Forschungsgemeinschaft through Grant No. EXC 153 (Excellence Cluster Universe") and by the European Union through the Initial Training Network Invisibles," Grant No. PITN-GA-2011-289442.

%

\end{document}